\documentclass[]{aastex7}

\usepackage{mathptmx}
\usepackage{graphicx}
\usepackage{dcolumn}
\usepackage{wrapfig}
\usepackage{amsmath}
\usepackage{multirow}
\usepackage{bm}
\usepackage{textcomp, gensymb}
\usepackage{mathrsfs}
\usepackage{xspace}
\usepackage{float}
\usepackage{upgreek}
\usepackage{enumitem}
\usepackage{pifont}     

\usepackage{tikz}
\usetikzlibrary{positioning,shapes,shadows,arrows,backgrounds,curvilinear,intersections,fit,matrix}
\usetikzlibrary{plotmarks,patterns}
\usepackage{pgfplots}
\usepackage{pgfplotstable}
\usepackage{booktabs}

\definecolor{darkred}{rgb}{0.6,0,0}
\definecolor{firebrick}{rgb}{0.75,0.125,0.125}
\definecolor{darkorange}{rgb}{1.0, 0.22, 0.}
\definecolor{darkgreen}{rgb}{0,0.5,0}
\definecolor{lightgray}{gray}{0.9}
\definecolor{gray}{rgb}{0.4,0.4,0.4}

\hypersetup{
  pdftitle = {Heavy Metal},
  pdfcreator = {LaTeX, pdfTeX},
  colorlinks = true,
  plainpages = false,
  linkcolor = firebrick,
  citecolor = firebrick,
  urlcolor = darkorange,
  bookmarksopen = false,
  linktocpage = true
}

\newcommand{\sib}[1]{\textsc{Sibyll}\,#1\xspace}
\newcommand{\qgsii}{\textsc{QGSJet~II-04}\xspace}
\newcommand{\eposlhc}{\textsc{Epos-LHC}\xspace}

\newcommand{\avg}[1]{\ensuremath{\langle{#1}\rangle}}

\newcommand{\Xmax}{\ensuremath{X_\text{max}}\xspace}

\newcommand{\DeltaXmax}{\ensuremath{\Delta\Xmax}\xspace}

\newcommand{\erange}[2]{\ensuremath{10^{#1}\text{ to }10^{#2}\,\text{eV}}\xspace}

\newcommand{\EeV}{\text{EeV}\xspace}
\newcommand{\eV}{\text{eV}\xspace}
\newcommand{\gcm}{\text{g}/\text{cm}^{2}}

\newcommand{\change}[1]{#1}

\begin{document}

\title{A Heavy-Metal Scenario of Ultra-High-Energy Cosmic Rays}

\author[orcid=0000-0002-7945-3605]{Jakub Vícha}
\affiliation{Institute of Physics of the Czech Academy of Sciences, Prague, Czech Republic}
\email[show]{vicha@fzu.cz}
\correspondingauthor{Jakub Vícha}
\author[orcid=0000-0003-1001-4484]{Alena Bakalová}
\affiliation{Institute of Physics of the Czech Academy of Sciences, Prague, Czech Republic}
\email{bakalova@fzu.cz}
\author[orcid=0000-0002-8473-695X]{Ana L. Müller}
\affiliation{Institute of Physics of the Czech Academy of Sciences, Prague, Czech Republic}
\email{mulleral@fzu.cz}
\author[orcid=0000-0001-6393-7851]{Olena Tkachenko}
\affiliation{Institute of Physics of the Czech Academy of Sciences, Prague, Czech Republic}
\email{tkachenko@fzu.cz}
\author[orcid=0000-0002-7943-6012]{Maximilian K. Stadelmaier}
\affiliation{Università degli Studi di Milano, Dipartimento di Fisica \& INFN, Sezione di Milano, Milano, Italy}
\affiliation{Karlsruhe Institute of Technology, Institut für Astroteilchenphysik, Karlsruhe, Germany}
\email{max.stadelmaier@posteo.net}

\begin{abstract}

%
The mass composition of ultra-high-energy cosmic rays is an open problem in astroparticle physics.
It is usually inferred from the depth of the shower maximum (\Xmax) of cosmic-ray showers, which is only ambiguously determined by modern hadronic interaction models.
We examine a data-driven scenario, in which we consider the expectation value of \Xmax as a free parameter.
We test the novel hypothesis whether the cosmic-ray data from the Pierre Auger Observatory can be interpreted in a consistent picture, under the assumption that the mass composition of cosmic rays at the highest energies is dominated by high metallicity, resulting in pure iron nuclei at energies above $\approx40\,\EeV$.
We investigate the implications on astrophysical observations and hadronic interactions, and we discuss the global consistency of the data assuming this heavy-metal scenario.
We conclude that the data from the Pierre Auger Observatory can be interpreted consistently if the expectation values for \Xmax from modern hadronic interaction models are shifted to larger values.

%

\end{abstract}



\section{Introduction}
\label{sec:intro}

The mass composition of ultra-high-energy cosmic rays (above $10^{18}\,\eV$) is an important puzzle piece concerning the question of the origin of the most energetic particles in the Universe.
These rare particles produce extensive air showers of secondary particles that are detected by large observatories, such as the Pierre Auger Observatory \citep{PierreAuger:2015eyc} and Telescope Array \citep{TAproject}.
The observables that are the most commonly used as mass estimators of the primary cosmic-ray particles are the depth of the shower maximum (\Xmax) and the number of muons produced during the development of the air shower and reaching the ground. 
The value of \Xmax is measured from the top of the atmosphere in units of g/cm$^{2}$ and indicates where the air shower reaches its maximum particle count.

Another important observable in the context of air-shower development is the number of muons created in the cascade.
An accurate estimate of the number of muons that reach the ground is very difficult to achieve using modern air-shower models; this pathology is known as the muon problem~\citep{WHISParticle}.
The state-of-the-art hadronic interaction models do not agree on the average \Xmax, nor on the number of hadronic shower particles (pions, muons, etc.) produced in air showers of the same type of primary particle and primary energy. 
These model disagreements are a consequence of conceptual differences in extrapolations of hadronic interaction properties studied at much lower energies and different kinematic regions using human-made accelerators.
However, these models produce consistent and stable expectations for fluctuations in both observables \citep{MuonFluct2020}. 

With the advent of novel machine-learning techniques being applied to the Pierre Auger Observatory (Auger) Surface Detector data~\change{\citep{PierreAuger:2024flk,AugerDNN:PRD}}, an unprecedentedly precise estimation of the mean and fluctuation of \Xmax was recently achieved for primary energies ranging from $3\,\EeV$ up to $100\,\EeV$.
However, the resulting moments of the logarithmic atomic mass number ($\ln A$) cannot be consistently interpreted using modern hadronic interaction models.
In case of the \qgsii~\citep{Qgsjet} model of hadronic interactions, a negative and thus nonphysical variance of logarithmic atomic mass $\ln A$, $\sigma^2(\ln A)$, is obtained even when accounting for systematic uncertainties.
In the case of the \sib{2.3d}~\citep{Sibyll} and \eposlhc~\citep{EposLHC} models, the predictions lie at the edge of the systematic uncertainty range of the data.
The variance $\sigma^2(\ln A)\approx0$ suggests a pure beam of primary cosmic rays above $E\simeq5\,\EeV$ (within uncertainties), while the energy evolution of the average \Xmax is only consistent with a gradual increase in the average logarithmic mass \avg{\ln A} as a function of the primary energy.
These inconsistencies could be resolved for all three models if the $\Xmax$ scale of models is shifted towards deeper values for the same primary particles.

Recently, fits of two-dimensional distributions of the ground signal and \Xmax in the energy range $3\,\EeV-10\,\EeV$ have revealed an alleviation of the muon problem~\citep{PierreAuger:2024neu} reducing it by approximately half compared to the previous analyses in~\citet{AmigaMuons,InclinedMuons,TestingHadronicInteractions}.
To match the two-dimensional distributions with a consistent interpretation of the mass composition, the expected scale of the number of hadronic shower particles needs to be increased by $\approx15\,\%-25\,\%$. 
At the same time, the simulated average \Xmax from all three aforementioned models \change{tends} to be shifted independently of the primary mass by $\approx20\,\gcm-50\,\gcm$, \change{implicitly, for a heavier mass composition than for the unmodified model predictions.}
\change{Such primary-mass independent shifts of the predicted \Xmax scale} could be qualitatively explained by a change in the normalization of the energy evolution of elasticity or multiplicity in hadronic interactions \citep{PierreAuger:2024neu}.


The \Xmax fluctuations produced for simulated showers from iron nuclei are too low using the \eposlhc model, being at the level of total defragmentation of the Fe nucleus.
This issue will be resolved in the upcoming model version \textsc{Epos-LHCR}~\citep{Pierog:2023ahq}, which will produce \Xmax fluctuations at the level of expectations from the \qgsii and \sib{2.3d} models.
Therefore, in this work, we consider the adjusted mass-composition model only for \sib{2.3d} and \qgsii, as the full implementation of \textsc{Epos-LHCR} was not yet available for common air-shower simulation frameworks.
Simulations using an upcoming \textsc{QGSJet~III} model~\citep{QGSJetIII} increase the predicted \Xmax scale by $\approx5$~g/cm$^2$ only, leaving, however, the \Xmax fluctuations approximately unchanged.
Finally, modifying properties of hadronic interactions of \sib{2.3d} \citep{MochiICRC23}, the universality of \Xmax fluctuations for iron nuclei was confirmed to be well within $\approx2~\gcm$.

In this heavy-metal scenario, first introduced in \citet{HeavyMetalUHECR2024}, we discuss a novel view on the mass-composition data of the Pierre Auger Observatory using two simple premises. Firstly, we consider the possibility of shifting the \Xmax scale predicted by models of hadronic interactions by the same energy-independent value for all primary species while keeping all other predictions of hadronic interaction models unchanged. 
Secondly, we postulate that the data agrees with expectations for pure iron nuclei above $10^{19.6}\,\eV$ ($\approx40\,$E\eV). 
While other studies investigate the potential presence of even heavier nuclei in the cosmic-ray flux at the highest energies \citep[see e.g.,][]{MetzgerHeavyNuclei, FarrarBNS, ZhangUHUHECR}, \change{triggered also by recent observation of one of the most energetic particle without clear association of a potential source direction \citep{MostEnergeticEventAtTA,UF_Amaterasu},} our assumption of pure iron nuclei is based on its status as the most abundant heavy element in the Universe and the heaviest nucleus produced in significant quantities through stellar nucleosynthesis.
Our two assumptions result in a generally heavier mass composition in a broad range of energies than what is obtained from the standard analyses of the $\Xmax$ moments~\citep{PierreAuger:2023kjt} and from the analyses that combine mass-composition and spectrum information~\citep{PierreAuger:2016use}, while resolving some tension of the muon puzzle.

The article is built up as follows:
the next section, Section~\ref{sec:XmaxScale}, describes the derivation of the shift in the predicted \Xmax scale from the most precise available data on \avg{\Xmax}.
We then present the new mass-composition scenario of ultra-high-energy cosmic rays in Section~\ref{sec:PrimFractions}, using four primary species by fitting the measured \Xmax distributions to the shifted model predictions on \Xmax.
The estimation of the consequences of this new scenario on energy spectra of individual cosmic-ray species is given in Section~\ref{sec:EnergySpectrum}, studies of hadronic interactions in Section~\ref{sec:HadronicInteractions}, and effect on backtracking of arrival directions in the Galactic magnetic field is presented in Section~\ref{sec:ArrivalDirections}.
We finally discuss the consistency with other studies and a possible presence of iron nuclei in cosmic-ray sources in Section~\ref{sec:Discussion}.

   \begin{figure*}
    \includegraphics[width=0.46\textwidth]{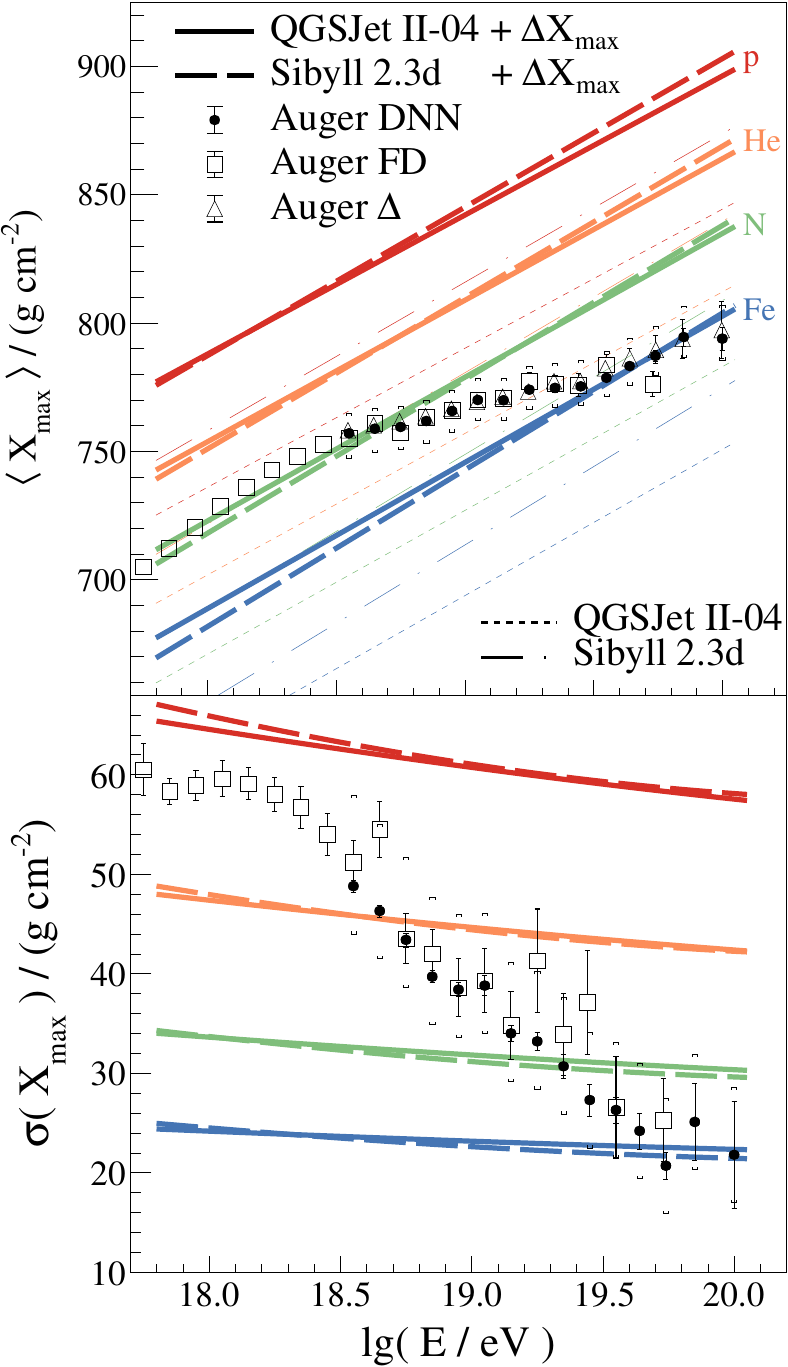}
    \hspace{1cm}
    \includegraphics[width=0.45\textwidth]{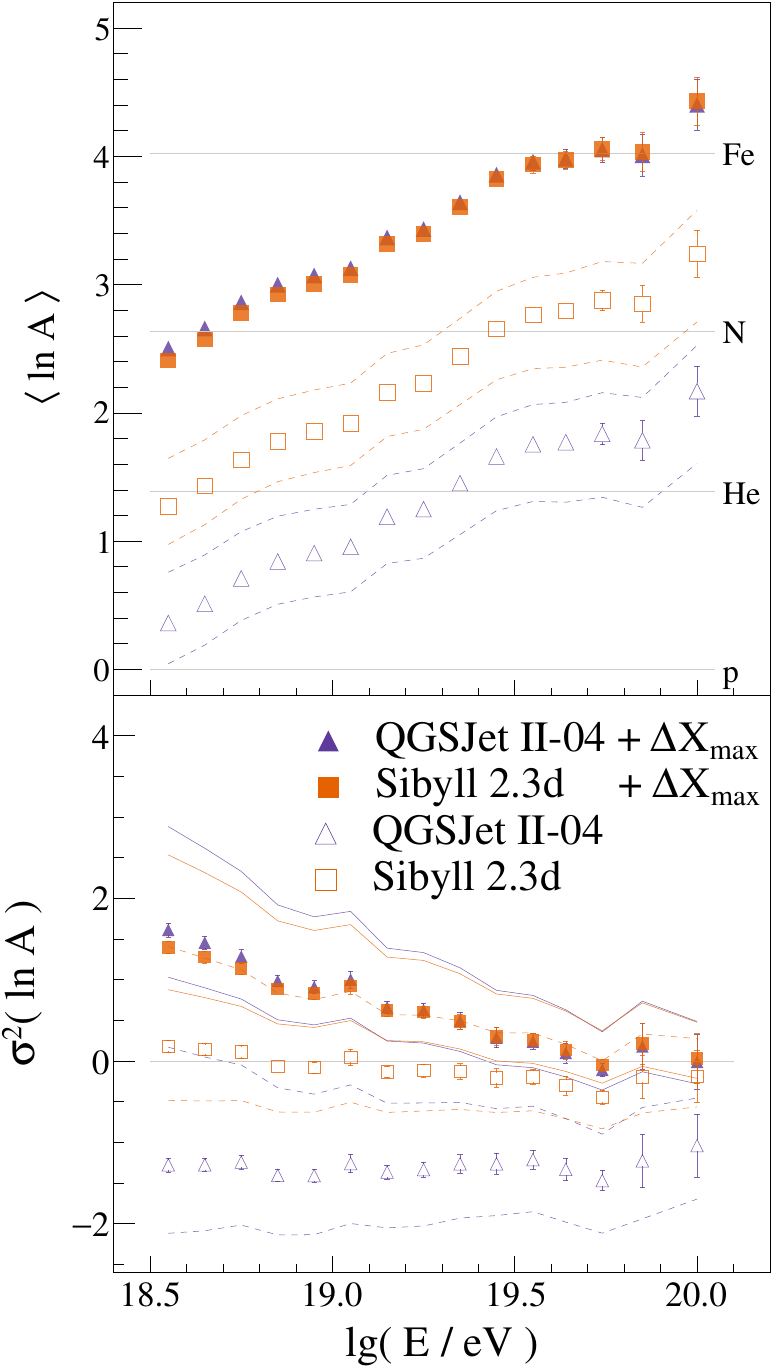}
    \caption{Left panel: The energy evolution of the mean and standard deviation of \Xmax for data \citep{PierreAuger:2024flk,XmaxICRC19,DeltaICRC19} measured by the Pierre Auger Observatory (black), including the systematic uncertainty in case of Auger DNN (brackets). The original model predictions for four primary species are depicted by thin lines, and the adjusted predictions by thick lines. Right panel: The energy evolution of the two lowest moments of $\ln A$, interpreted from the Auger DNN data shown in the left panel, using the adjusted (full markers) and original (open markers) model predictions.}
    \label{fig:XmaxScaleFit}
  \end{figure*}

\begin{figure*}
    \includegraphics[width=1.\textwidth]{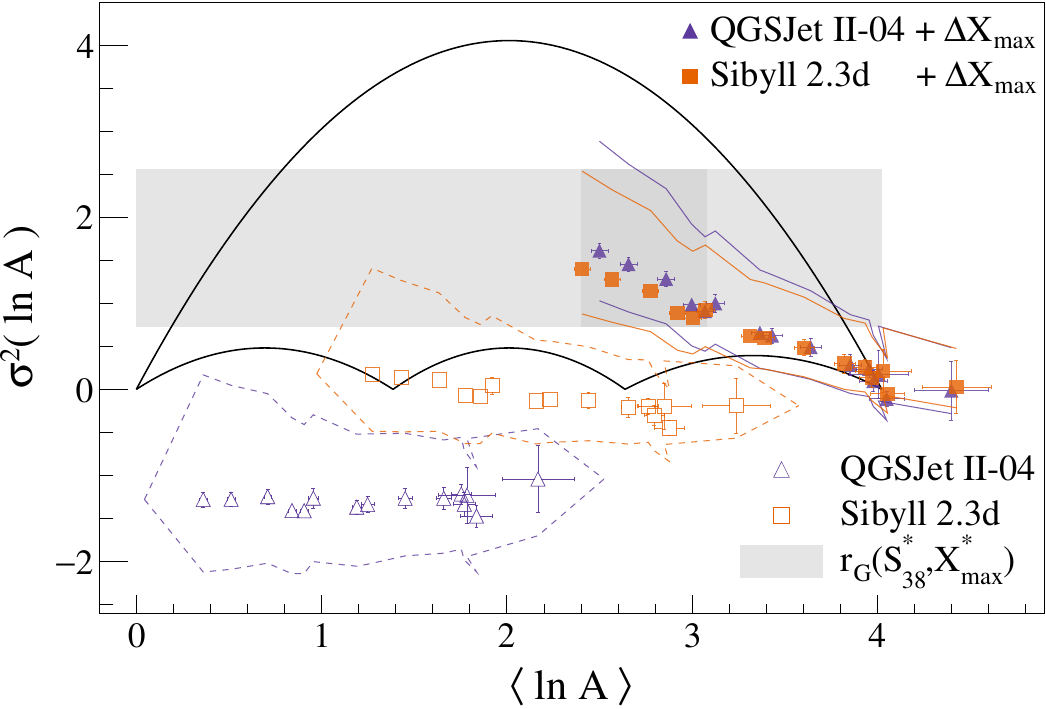}
    \caption{The relation between the interpreted moments of $\ln A$ with (full markers) and without (open markers) the application of \DeltaXmax to the model predictions using Auger DNN data~\citep{PierreAuger:2024flk}. The dashed and solid lines correspond to the systematic uncertainties. The range of possible combinations for p, He, N and Fe nuclei is limited by the black lines forming the umbrella. The resulting value of $\sigma^2(\ln A)$ inferred from the correlation coefficient $r_{G}$ between the ground signal and \Xmax \citep{MixedAnkle} for $3\,\EeV-10\,\EeV$ is shown by the gray band, with the darker and narrower band indicating the respective energy range of the points for \DeltaXmax.
    }
    \label{fig:Umbrella}
\end{figure*}

\section{Adjustment of Predicted \Xmax Scale}
\label{sec:XmaxScale}

The mean and standard deviation of the \Xmax distribution, as estimated from direct Fluorescence Detector measurements~\citep{XmaxICRC19}, as well as from the Surface Detector data, are depicted in the left panel of Fig.~\ref{fig:XmaxScaleFit}. 
These are presented alongside both the shifted and standard reference values of the \qgsii and \sib{2.3d} models.
The Surface-Detector data is interpreted using two methods: a Deep Neural Network (DNN) method~\citep{PierreAuger:2024flk} and the $\Delta$-method, which relies on a detector time-response parameter empirically proven to be correlated with $\Xmax$~\citep{DeltaICRC19}.
It is important to note that both Surface-Detector estimates are calibrated to match the Fluorescence-Detector data, applying a positive shift of about $30\,\gcm$ in the predicted $\Xmax$, independently of this work.

The fluctuations of \Xmax above $10^{19.6}\,\eV$ are consistent with expectations for pure iron nuclei, as predicted by the two models, with p($\chi^{2})>0.5$.
Moreover, in the same energy range, the mean \Xmax follows the constant elongation rate predicted for a single type of primary particles, indicating the possibility of consistently describing both \Xmax moments by a pure beam of particles. 
The $\chi^{2}$ tests, given in the Supplementary material (Fig.~\ref{fig:BeamPurity}), also include a scenario using silicon as the only primary particle type.
In such a pure-silicon scenario, the measured \Xmax fluctuations are well described only above $10^{19.8}\,\eV$, where the event statistics are very low.

The shift \DeltaXmax for the \avg{\Xmax} values is obtained by assuming pure iron nuclei above $10^{19.6}\,\eV$ and maintaining the elongation rate ($\text{d}\Xmax / \text{d}\lg \text{E}$) predicted by the hadronic interaction models.
In this way, we fit the \avg{\Xmax} from the Auger DNN data obtaining $\DeltaXmax=52\pm1^{+11}_{-8}\,\gcm$ and $\DeltaXmax=29\pm1^{+12}_{-7}\,\gcm$ for \qgsii and \sib{2.3d}, respectively.
These values are consistent with the results from \citet{PierreAuger:2024neu} at the statistical level, which is not valid in the case of the pure-silicon scenario (see the Supplementary material~\ref{app:PurityTests}).

The energy evolutions of the mean and variance of $\ln A$ of the cosmic rays, interpreted (see~\cite{InterpretationOfXmax}) from the \Xmax distributions, are depicted in the right panel of Fig.~\ref{fig:XmaxScaleFit} alongside the interpretations using the unmodified model predictions.
Note that no systematic-uncertainty band is shown for \avg{\ln A} in the case of \DeltaXmax, since \avg{\ln A} is always fitted to the expectation for iron nuclei above $10^{19.6}\,\eV$, independently of the absolute values of measured \Xmax.
Applying the shift \DeltaXmax, the energy evolutions of $\avg{\ln A}$ and $\sigma^{2}(\ln A)$ consistently describe an increase in the average mass of cosmic rays as well as a decrease in the width of the mix of primary masses; as required, we arrive at a pure beam of iron nuclei at the highest energies.

Besides the overall mass shift, the $\ln A$ moments, as derived from the shifted $\Xmax$ scales of the four primary particles, yield a far-reaching difference: the updated moments can be interpreted consistently within model-independent expectations.
The $\ln A$ moments (before and after applying the shift in $\Xmax$) are depicted in Fig.~\ref{fig:Umbrella}, alongside the allowed region for all possibilities when mixing four different primary species (protons, and helium, nitrogen and iron nuclei). 
The moments of $\ln A$ as interpreted according to \sib{2.3d} and \qgsii both lie (largely) outside this \emph{umbrella} shape. 
The two $\ln A$ moments obtained after the $\DeltaXmax$ shift are within the allowed region.
Furthermore, the updated moments are now consistent with the results from the model-independent correlation of the depth of the shower maximum with the size of the shower footprint between $3\,\EeV$ and $10\,\EeV$~\citep{MixedAnkle}, as indicated in Fig.~\ref{fig:Umbrella} by dark-shaded box for the corresponding energy range of \DeltaXmax points.

\section{Energy evolution of relative primary masses}
  \label{sec:PrimFractions}

To estimate the energy evolution of the relative abundance of different primary-mass species in the heavy-metal scenario, we use the \Xmax distributions published by~\citet{Auger-LongXmaxPaper} to fit four primary fractions (protons, as well as He, N, and Fe nuclei) to the expectations from the \qgsii and \sib{2.3d} models, after accounting for \DeltaXmax. 
The \Xmax distribution templates used in the fit were generated using the CONEX air shower simulation code~\citep{CONEX}, with the same energy distribution per bin as in~\cite{Auger-LongXmaxPaperMass}, and were modified to account for the effects of detector acceptance and resolution accordingly. 
To increase the event statistics, we merged adjacent energy bins above $10^{18.4}\,\eV$, reducing statistical fluctuations in the fitted primary fractions. 
The last energy bin is integral and includes all events with energies above $10^{19.5}\,\eV$. 
The fitted primary fractions are shown in the top panels of Fig.~\ref{fig:mass_spectrum}, and the fitted \Xmax distributions are provided in the Supplementary material~\ref{app:XmaxMassFits}, along with the corresponding $p$-values. 
Overall, with \sib{2.3d}, we achieve sufficiently good fit qualities ($p$-values $\geq$ 0.05) across the entire energy range from \erange{17.8}{20}. 
The \qgsii model provides an acceptable fit above $10^{18.3}\,\eV$, but performs poorly below this energy, with $p$-values less than $10^{-4}$.

Both hadronic interaction models favor a mass composition dominated by iron nuclei across all energies, with the iron fraction reaching a minimum between $30\%$ and $50\%$ at $3\,\EeV-4\,\EeV$, followed by a rise to nearly 100\% at the highest energies \change{as assumed}. 
The energy evolution of the estimated mass composition is consistent with the transition from lighter to heavier components observed in \cite{Auger-LongXmaxPaperMass}, but with a smaller presence of lighter nuclei in the mix. 
The contributions of helium nuclei and protons individually remain below $\approx30\%$ at the lowest energies and decrease to below $\approx10\%$ above the ankle feature of the cosmic-ray energy spectrum ($\approx5\,\EeV$) . 
Above this point, around $\approx10\,\EeV$, the nitrogen fraction rises to about $30\%-50\%$ before also dropping to nearly zero at the end of the cosmic-ray spectrum.

  \begin{figure*}
    \includegraphics[width=0.44\textwidth]{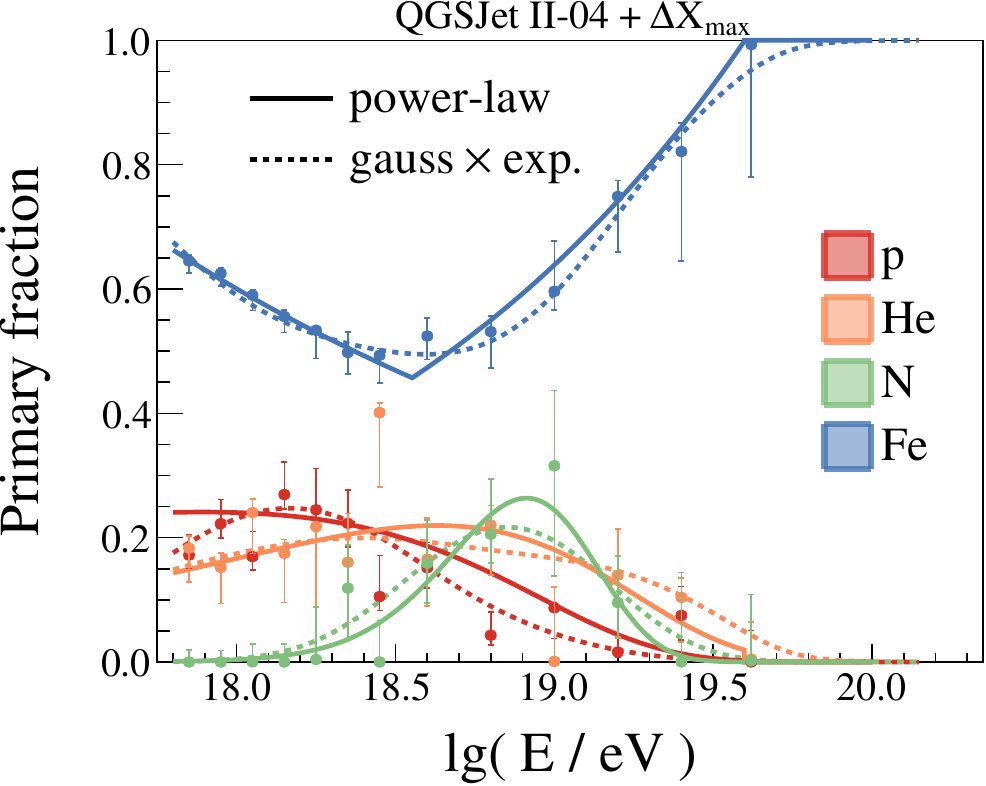}\hfill
    \hfill
    \includegraphics[width=0.44\textwidth]{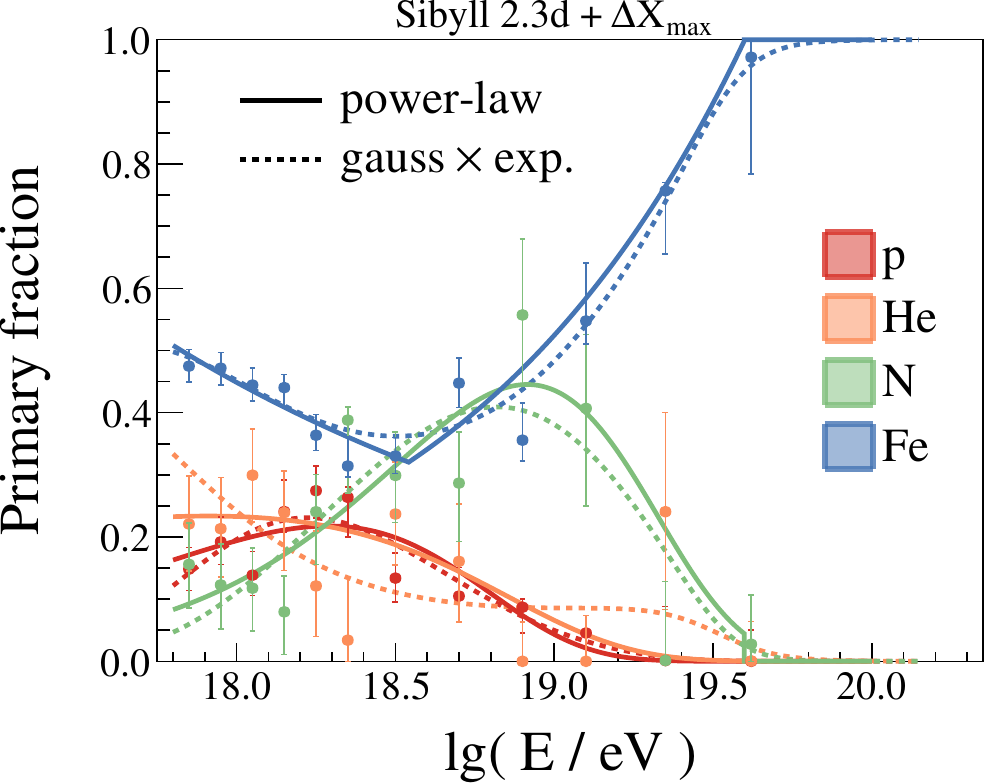}\\

    \includegraphics[width=0.44\textwidth]{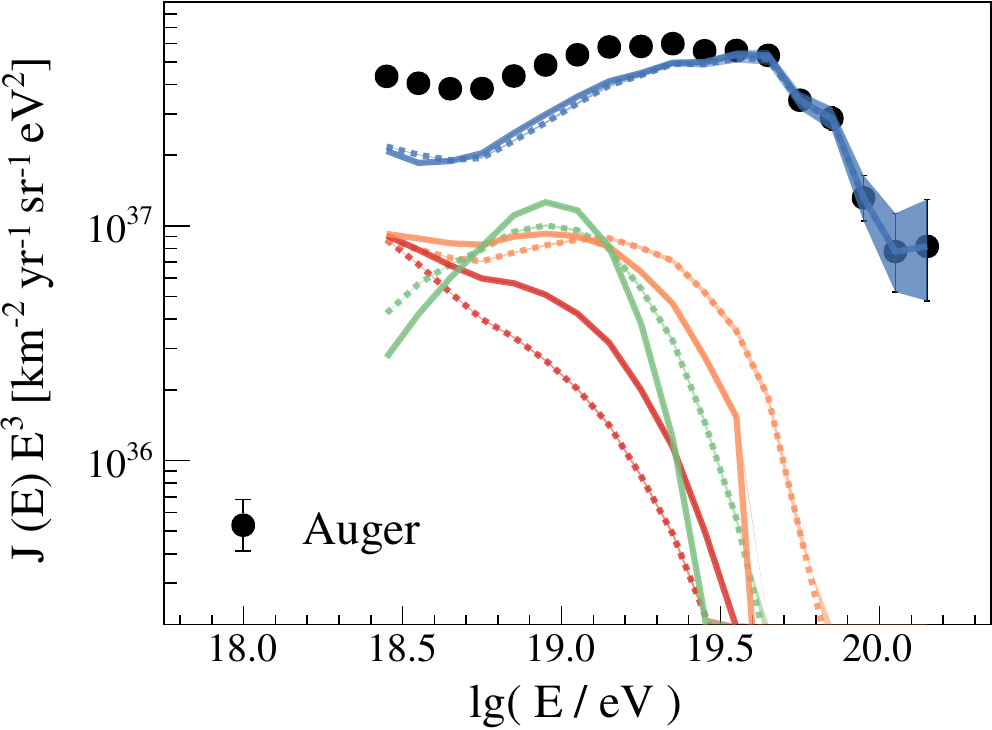}
    \hfill
    \includegraphics[width=0.44\textwidth]{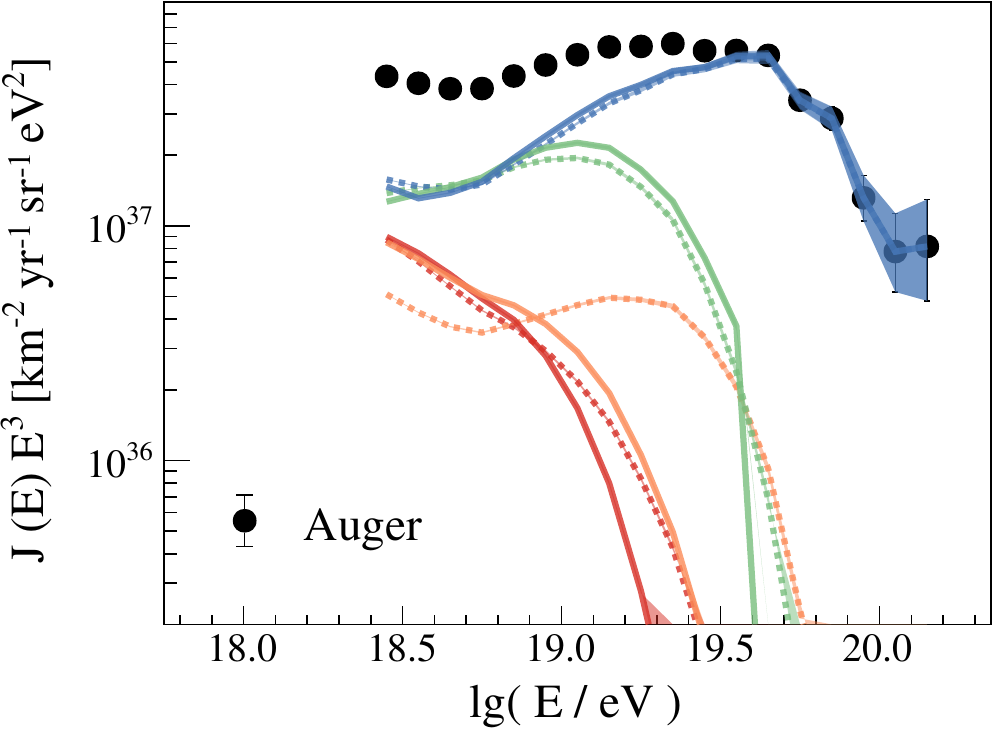}\\
    
    \includegraphics[width=0.44\textwidth]{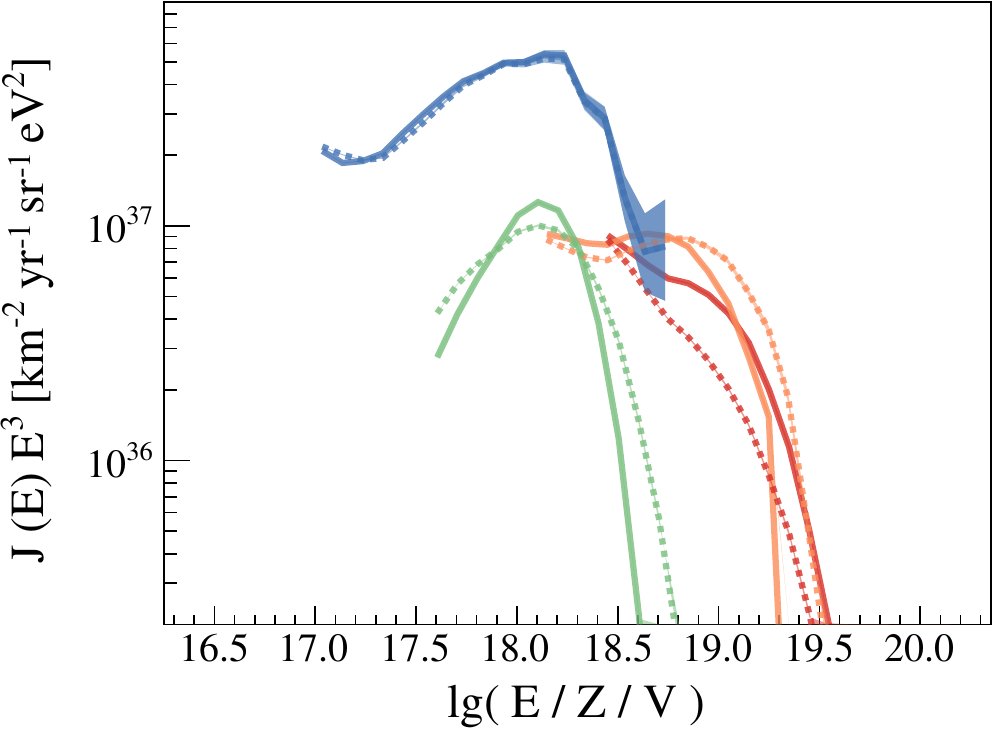}
    \hfill
    \includegraphics[width=0.44\textwidth]{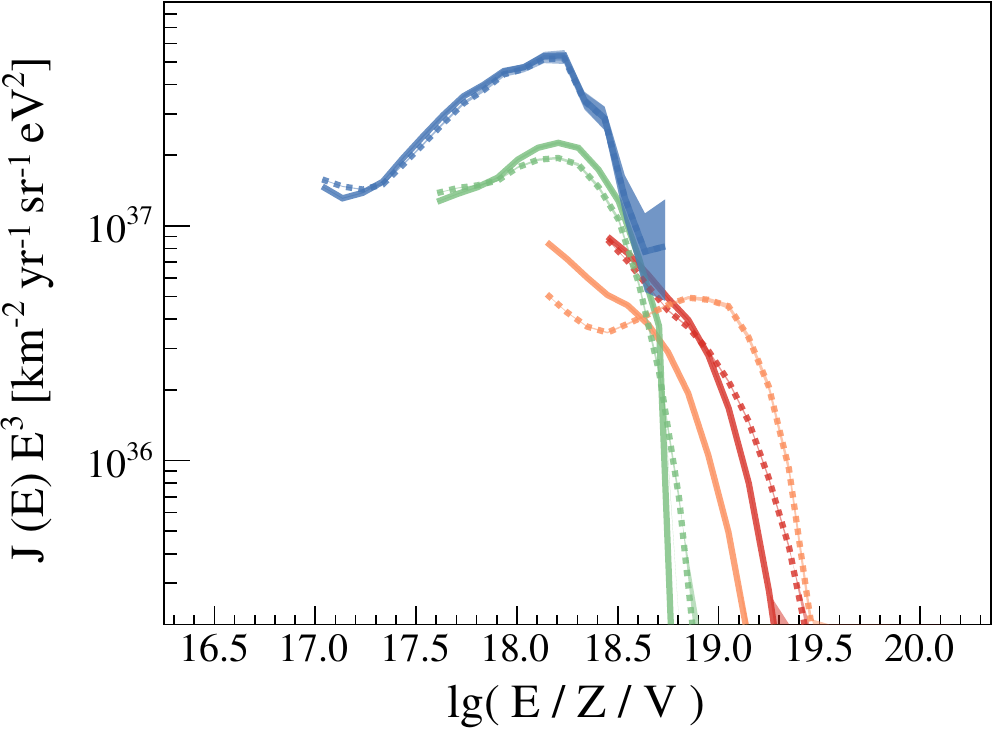}
    \caption{Energy evolutions of the primary fractions (top panels) and evolutions of differential fluxes of individual primary species as functions of energy (middle panels) and rigidity (bottom panels) using \qgsii + \DeltaXmax (left panels) and \sib{2.3d} + \DeltaXmax (right panels). The primary fractions were obtained by fitting the Auger \Xmax distributions~\citep{Auger-LongXmaxPaper}, while the total differential flux of cosmic rays was taken from ~\citet{SDEnergySpectrum2020}. The two parameterized energy evolutions of primary fractions, see Section~\ref{sec:PrimFractions}, are represented using dashed and solid lines in the panels.}
    \label{fig:mass_spectrum}
  \end{figure*}

We used two functional forms to describe the energy evolution of fitted primary fractions: a smoothed Gaussian multiplied by an exponential function and power-law functions with a simple exponential cutoff; the latter being mentioned in \citet{PierreAuger:2016use}. 
The functional forms and parameters for both parameterizations are given in the Supplementary material~\ref{app:functions_fractions}.
In this work, we use exclusively public data of the Pierre Auger Observatory. 
However, we note that including additional data will enable a more precise estimation of the primary fractions, even within the heavy-metal scenario, leading to a more accurate determination of how individual primary fractions evolve with energy.

\section{Energy Spectrum of Individual Primaries}
\label{sec:EnergySpectrum}

Using the primary fractions found in Section~\ref{sec:PrimFractions}, we decompose the energy spectrum of cosmic rays at the highest energies into contributions from different primary nuclei.
The all-particle energy spectrum\footnote{Note that for better visibility, the flux is scaled by the energy to the power of three.} of cosmic rays, as measured by the Pierre Auger Observatory~\citep{SDEnergySpectrum2020}, is shown in the middle panels of Fig.~\ref{fig:mass_spectrum}.
The individual contributions of the four primary species, according to the two parameterizations of primary fractions shown in the top panels, are depicted alongside the total spectrum in the respective colors.
From this decomposed energy spectrum, the rigidity ($E/Z$) of each mass component can be inferred.
The rigidity dependence of cosmic-ray flux is shown in the bottom panels of Fig.~\ref{fig:mass_spectrum} for the four primary species.

By assumption, the flux suppression above ${\approx}40\,\EeV$ is caused solely by the spectral feature of iron nuclei, which might be related to the propagation effects or the maximum rigidity available at a source.
Interestingly, the instep feature at ${\approx}15\,\EeV$ in the cosmic-ray energy spectrum \change{\citep{PRL2020spectrum}} corresponds to the fading of nitrogen nuclei from the cosmic rays above this energy.
The rigidity cutoffs of iron and nitrogen nuclei seem to coincide at approximately $10^{18.2}\,\text{V}$, suggesting a common origin for these elements in the heavy-metal scenario.
The ankle feature, meanwhile, might be connected to the fading of the light component (i.e., protons and helium nuclei) from the cosmic-ray population.
However, the main limitation for these interpretations remains the statistical uncertainty in the energy evolution of primary fractions, which is especially large for protons and helium nuclei. 
On the other hand, their rigidity distributions might indicate, for example, the presence of a separate population of sources with different metallicities. The constraints could be mitigated in the future by using a larger statistical sample of measured data to refine the fit of the primary fractions.

\section{Hadronic Interactions}
  \label{sec:HadronicInteractions}

The number of muons produced in extensive air showers from ultra-high-energy cosmic rays is an interesting proxy for the amount of hadronic multi-particle production occurring in the first interactions.
The mass composition is a crucial aspect when studying hadronic interactions, as it shifts the expectations in the number of muons (hadrons) produced in showers and may alleviate as well as aggravate the muon puzzle.
In the following, we briefly discuss the implications of the heavy-metal scenario on the scale of the predicted number of muons and combined effect of modified elasticity and cross-section on the tail of \Xmax distribution.


\subsection{The Number of Muons}
  \label{sec:MuonNumber}

\begin{figure*}
    \includegraphics[width=0.44\textwidth]{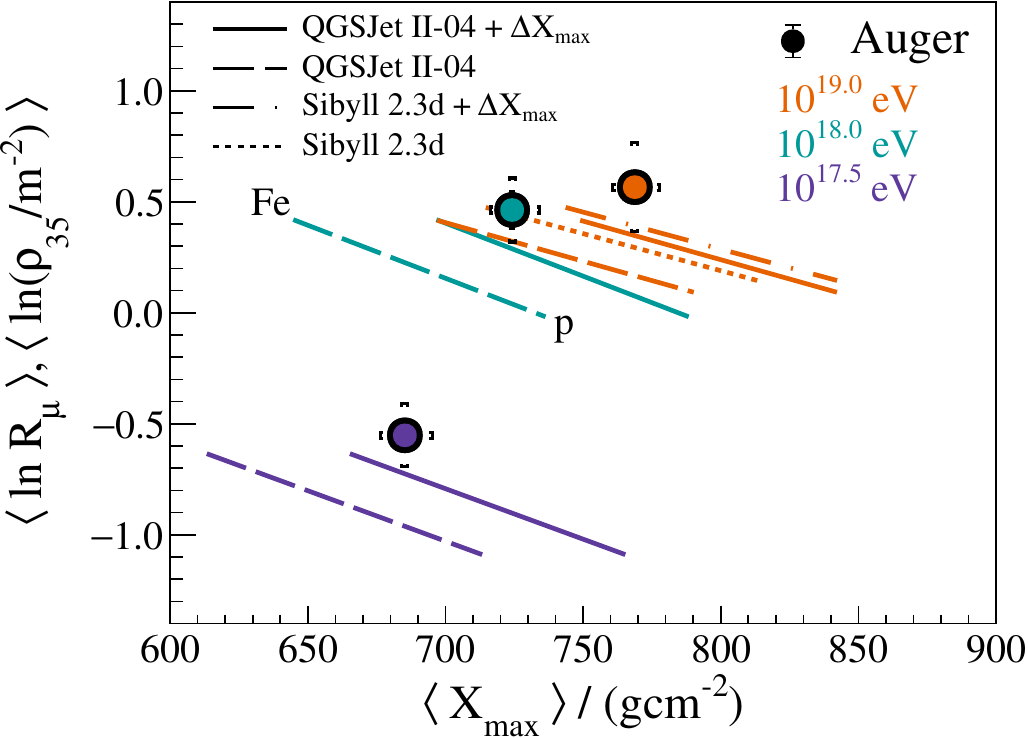}
    \hspace{1.3cm}
    \includegraphics[width=0.44\textwidth]{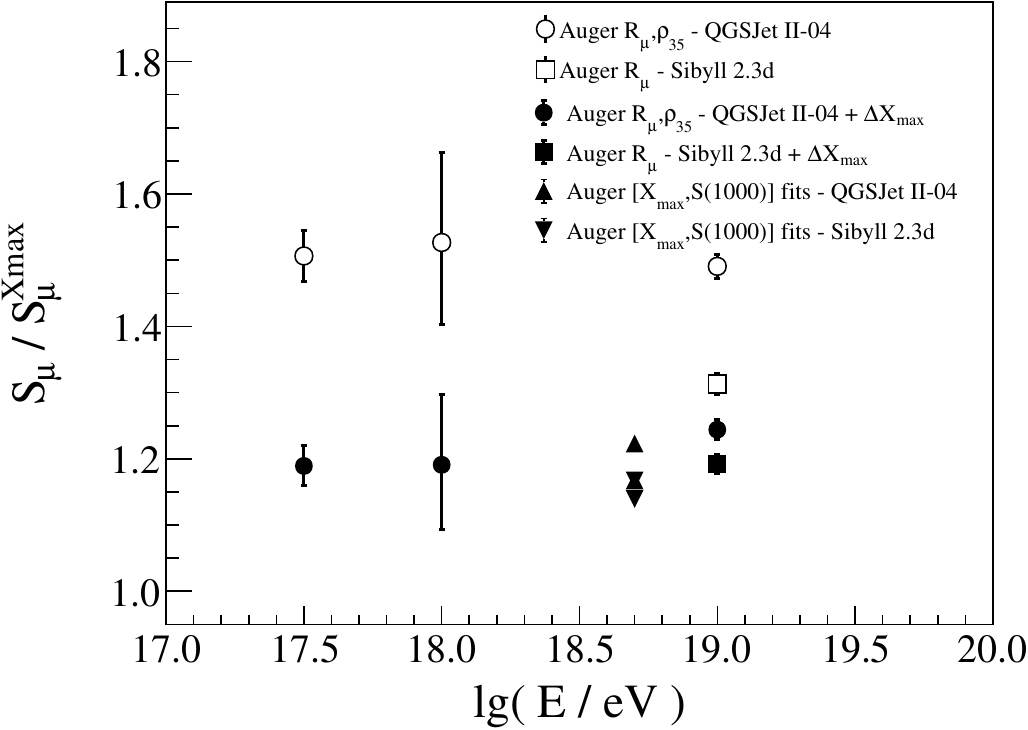} \\

    \includegraphics[width=0.45\textwidth]{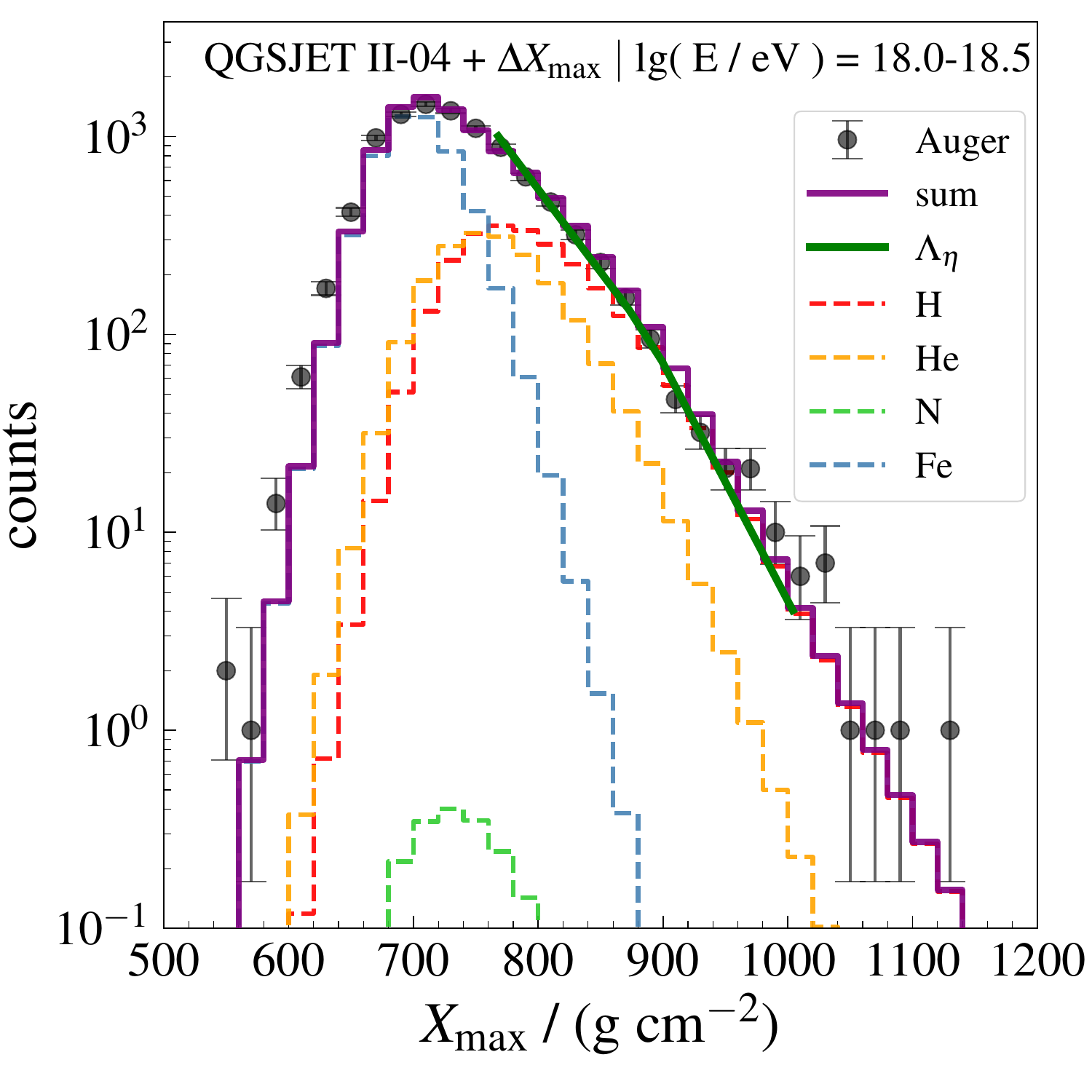}
    \hspace{1.25cm}
    \includegraphics[width=0.45\textwidth]{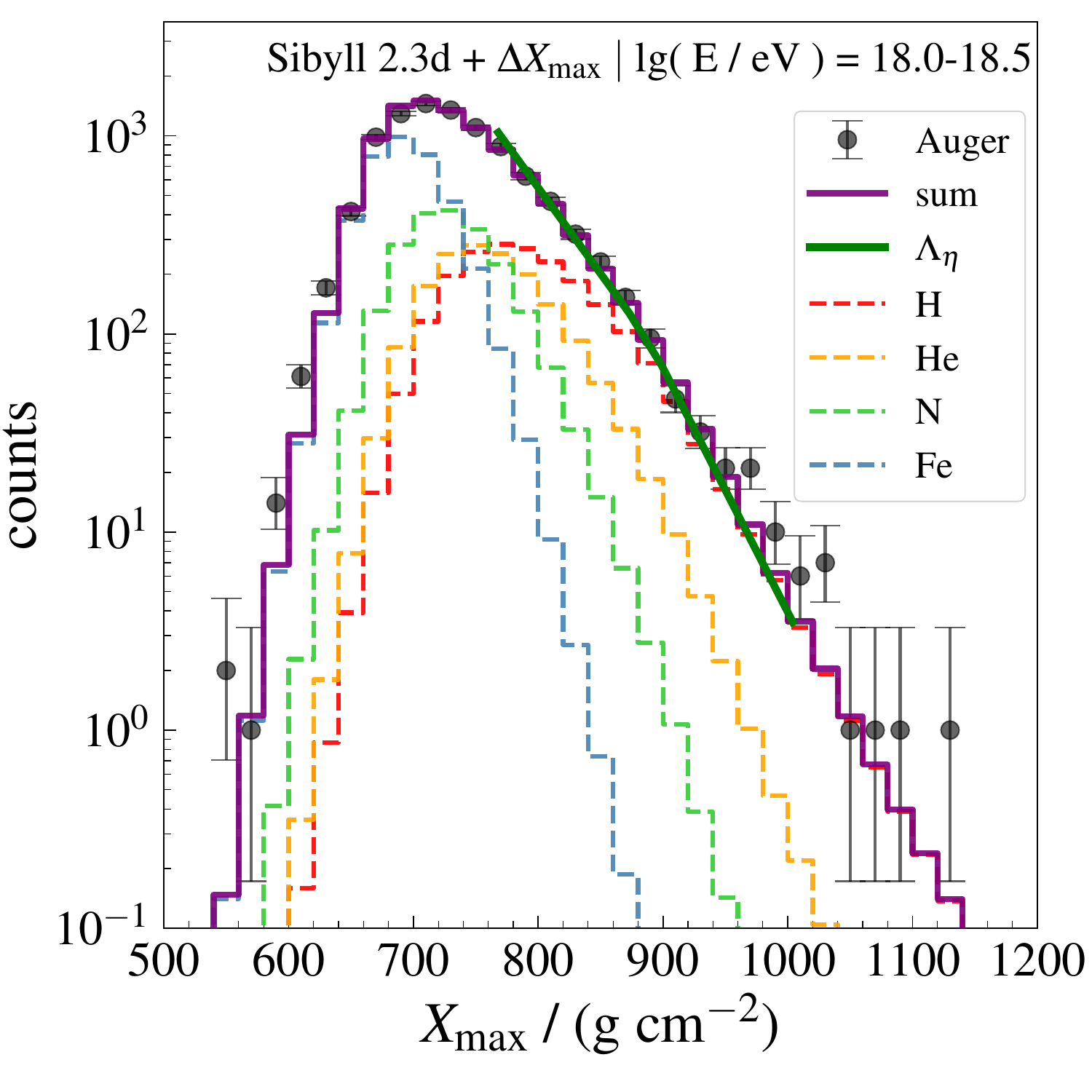}
    \caption{Top-left panel: The muon number obtained by direct measurements at Auger~\citep{AmigaMuons,MuonFluct2020} compared to the \change{predictions with and without the application of \DeltaXmax. Note that in case of the \sib{2.3d} model only the predictions at energy $10^{19}$\,\eV were at disposal}. Top-right panel: The ratio of the muon signal in data and the signal predicted from air shower simulations with (full markers) and without application of \DeltaXmax (open markers). The results from \cite{PierreAuger:2024neu} are displayed by triangles. Bottom panels: The \Xmax distribution of Auger data \citep{Auger-LongXmaxPaper} compared to the prediction for the heavy-metal scenario (gauss $\times$ exp.) using \qgsii+\DeltaXmax (left) and \sib{2.3d}+\DeltaXmax (right) in the energy bin $10^{18.0-18.5}\,\eV$. For visualization purposes, the fits to the tail of the predicted \Xmax distributions ($\Lambda_{\eta}$), corrected for detector acceptance, are shown.}

    \label{fig:Hadronics}
  \end{figure*}

In the top left panel of Fig.~\ref{fig:Hadronics}, we present the direct measurements of muon signal $S_{\mu}$ ($R_{\mu}$ \citep{InclinedMuons}, $\rho_{35}$ \citep{AmigaMuons}) by the Pierre Auger Observatory, along with the corresponding expectations from the \qgsii and \sib{2.3d} models \change{with and without applying \DeltaXmax, respectively.}
In the heavy-metal scenario, consistency with measured \avg{\Xmax} is achieved within the systematic uncertainty for the \sib{2.3d} model and is nearly within the systematic uncertainty in the case of the \qgsii model.

The muon deficit in the simulated data can be assessed from the ratio of the measured muon signal ($S_{\mu}$) to the expectation value from simulated showers with the same average depth of the shower maximum, \avg{\Xmax} ($S^{\Xmax}_{\mu}$).
Our results are depicted in the top right panel of Fig.~\ref{fig:Hadronics}.
We find that the muon deficit in the expectations from the \qgsii model is reduced from about 50\% to about 20\%-25\%.
In the case of \sib{2.3d} model, the lack of muons is reduced from about 30\% to approximately $20\%$.
Furthermore, the apparent muon deficit is independent of the energy of the primary particle and is consistent with the findings from~\citet{PierreAuger:2024neu} using the two-dimensional fits of the ground signal at 1000\,m, $S(1000)$, and \Xmax. 

\subsection{Tail of \Xmax distribution}
  \label{sec:CrossSection}
The tail of the \Xmax distribution provides essential insight into proton-air interactions, e.g., the inelastic p-air cross-section \citep{ProtonAirXsec-Auger}, as proton-induced showers penetrate deeper into the atmosphere before reaching their maximum development compared to air showers initiated by heavier nuclei. 
We investigate how well the measured tails of \Xmax distributions are described within the heavy-metal scenario.  
A fair description of the tail of the measured \Xmax distribution is shown in the bottom panels of Fig.~\ref{fig:Hadronics} by the shifted model predictions in the energy bin $10^{18.0-18.5}\,\eV$ for the \qgsii and \sib{2.3d} models.

We fit the \Xmax exponential shape ($\Lambda_{\eta}$) for the same \Xmax and energy range as in \citet{ProtonAirXsec-Auger}, for the four primaries according to our mass-composition model (gauss $\times$ exp.), applied to CONEX air-shower simulations smeared by the \Xmax resolution according to \citet{Auger-LongXmaxPaper}.
The resulting values, $\Lambda_{\eta}=(51.9\pm0.4)\,\gcm$ and $\Lambda_{\eta}=(50.0\pm0.4)\,\gcm$ for \qgsii$+\DeltaXmax$ and \sib{2.3d}$+\DeltaXmax$, respectively, are smaller than the value $\Lambda_{\eta}=\left(55.8\pm2.3(\rm stat)\pm1.6(\rm sys)\right)\,\gcm$ obtained from the Auger data \citep{ProtonAirXsec-Auger}.
\change{The smaller predicted values of $\Lambda_{\eta}$  in the heavy-metal scenario might be explained by a larger helium fraction, larger p-p inelastic cross-section, or lower elasticity extrapolated in the two models from accelerator measurements, which is qualitatively in line with the extrapolations used in the updated \eposlhc model \citep{Pierog:2023ahq} and predicted by studies of modifications of hadronic interactions \citep{MochiICRC23}.}
In the case of unmodified \qgsii and \sib{2.3d} \Xmax distributions, we obtain $\Lambda_{\eta}=(55.2\pm0.4)\,\gcm$ and $\Lambda_{\eta}=(52.1\pm0.4)\,\gcm$, respectively\change{, as a consequence of a lower helium fraction wrt. heavy-metal scenario}.
We emphasize that, in any case, our $\Lambda_{\eta}$ values are not directly comparable to those published by the Pierre Auger Collaboration, as all detector and event-selection effects have not been taken into account in our calculation. However, this approach represents the best possible comparison within our limitations.
Note that a more sophisticated method to derive the inelastic p-p cross-section is needed to accurately account for a mixed mass composition \citep{MassXsecICRC2021}.


\section{Arrival Directions}
  \label{sec:ArrivalDirections}


Cosmic rays are subject to magnetic deflections during their propagation from the source to the Earth.
The dipole anisotropy observed in the arrival directions of cosmic rays above 8~EeV suggests their extragalactic origin \change{\citep{PierreAuger:2017pzq, AugerDipole24}}.

In this section, we demonstrate the effect of the Galactic magnetic field (GMF) on the arrival directions by backtracking the cosmic rays as their anti-particles, assuming their mass according to the heavy-metal scenario. 
Firstly, we study the possible features of an extragalactic dipole in the distribution of cosmic-ray arrival directions, consistent with the observed dipole above $8$~EeV, after accounting for the effects of the GMF.
Secondly, we derive the region in the sky from which the most energetic events~\citep{AugerMostEnergeticEvents}, above $78$~EeV, might come, assuming they all are iron nuclei.

\begin{figure*}
    \centering
    \includegraphics[width=0.75\linewidth]{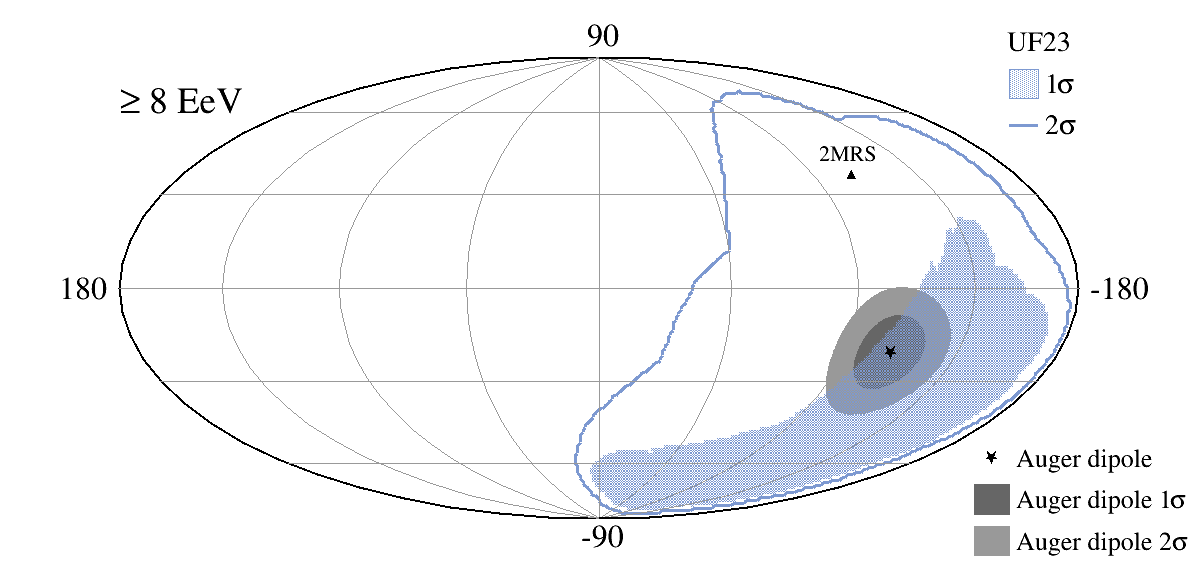}\\
    \hspace{1.5cm}\includegraphics[width=0.75\textwidth]{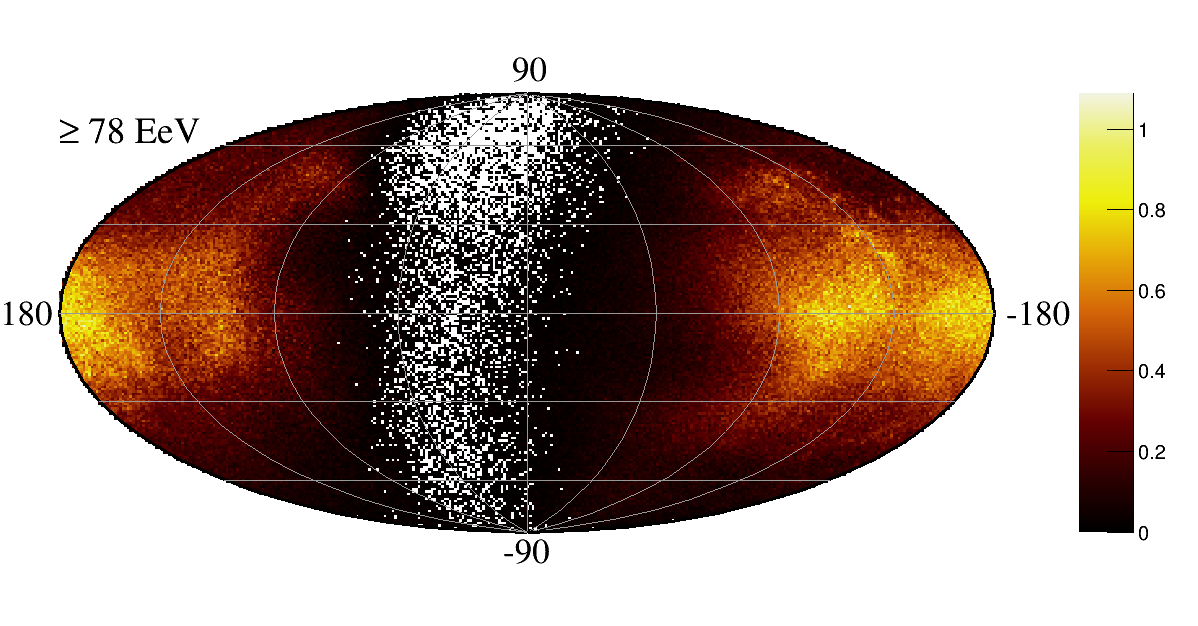}\\
    \hspace{1.5cm}\includegraphics[width=0.75\linewidth]{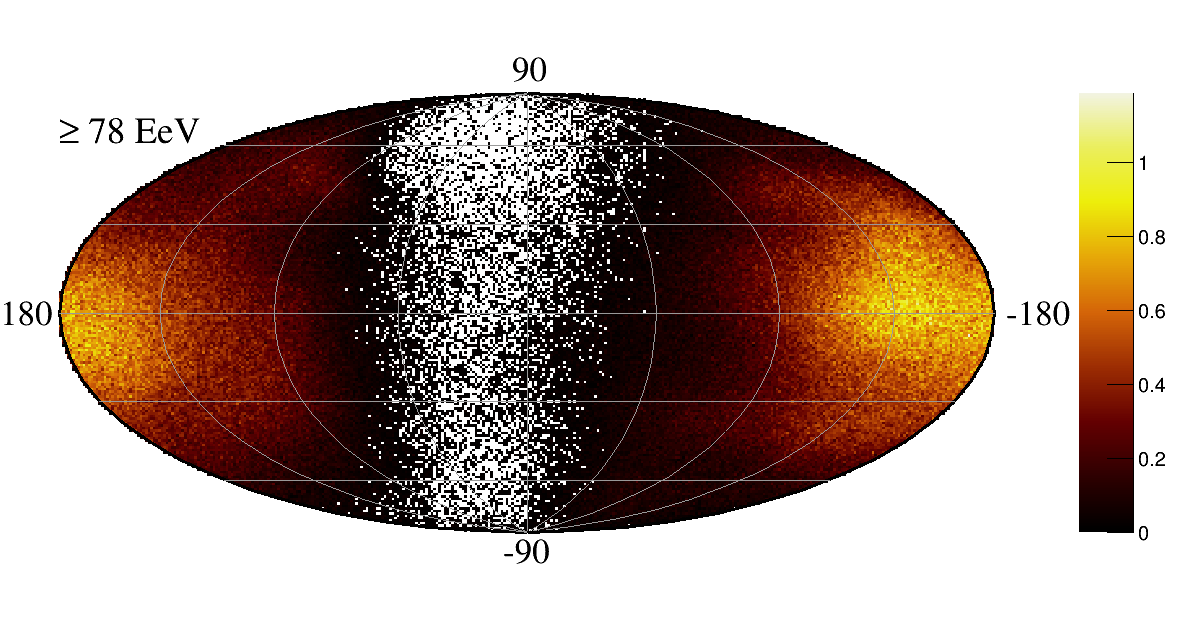}
    \caption{Top panel: Possible directions of an extragalactic dipole in the arrival directions of cosmic rays above 8\,\EeV, compatible at $1\sigma$ (blue area) and $2\sigma$ (blue contour) level with the Auger measurement \citep{PierreAuger:2017pzq}, for the UF23 models of the GMF, shown in Galactic coordinates. The mass composition scenario obtained for \sib{2.3d}+\DeltaXmax is used. Middle panel: The distribution of backtracked directions at the edge of the Galaxy for the 100 most energetic Auger events \citep{AugerMostEnergeticEvents}, shown in Galactic coordinates. Bottom panel: \change{Distribution of the backtracked directions at the edge of the Galaxy for isotropically distributed arrival directions of iron nuclei on Earth, shown in Galactic coordinates and accounting for the Auger geometrical exposure.}}
    \label{fig:ExtragalacticDipoleUF23}
\end{figure*}

\subsection{Features of an Extragalactic Dipole}
\label{sec:Dipole}

Assuming an ideal dipole distribution of the cosmic-ray flux at the edge of Galaxy\change{ (20\,kpc from the Galactic Center)}, we constrain the range of possible extragalactic features of the dipole by repeating the analysis from~\citet{BakalovaJCAP23}, using new models of the GMF and the mass-composition scenario for \sib{2.3d}+\DeltaXmax proposed in this work (see the dotted lines in the top right panel of Fig.~\ref{fig:mass_spectrum} and Supplementary material~\ref{sec:GaussExponentialForm}).  
The propagation of cosmic rays in the GMF is simulated with CRPropa~3~\citep{CRPropa32}. 
We use the eight models of the coherent component of the GMF from~\citet{UF23} (UF23), with the turbulent component from the model of the GMF by~\citet{JF12} (JF12), and apply the corrections from the Planck Collaboration \citep{JF12Planck}.  
The simulations were performed for three values of the coherence length of the turbulent component; $l_c=30$~pc, 60~pc and 100~pc, with multiple realizations of the turbulent field for each $l_c$.
The extragalactic direction of the dipole is imposed into all possible longitudes and latitudes with a step of $1^{\circ}$, and the extragalactic amplitude of the dipole $A_0$ is explored in the range from $6$\% up to $30$\%, in discrete steps of 2\%, and from $30$\% up to $100$\%, in discrete steps of $10$\%.  \change{We compare the reconstructed amplitude and the direction of the dipole on Earth from the simulations with the measured values by the Pierre Auger Observatory, an amplitude of $7.4^{+1.0}_{-0.8}$\% and direction in equatorial coordinates $(\alpha,\delta)=(97^{\circ}\pm8^{\circ},-38^{\circ}\pm9^{\circ})$ from \citet{AugerDipole24}}.

The identified possible extragalactic directions that are compatible with the direction and amplitude measurement from the Pierre Auger Observatory at the $1\sigma$ and $2\sigma$ levels are shown in the top panel of Fig.~\ref{fig:ExtragalacticDipoleUF23} for the combination of all eight UF23 models of the GMF (see Supplementary material~\ref{app:Dipole} for the solutions corresponding to each individual UF23 model \change{and the effect of different coherence lengths of the turbulent component}).
Note that possible extragalactic directions of the dipole at the $1\sigma$ level were identified only for amplitudes above 40\%.
All extragalactic dipoles identified in this work have an initial amplitude of $A_0\geq12\%$ (see Fig.~\ref{fig:DipoleAmplitudes} in the Supplementary material). 
The identified areas of possible extragalactic directions are consistent with the previously published results from~\citet{BakalovaJCAP23}, where no specific mass composition was assumed, and different GMF models were used. 
The direction of the 2 Micron All-Sky Redshift Survey (2MRS) dipole~\citep{2MRS}, which describes approximately the distribution of matter in the local Universe, is located within the identified region at the $2\sigma$ level. 
The large amplitudes of an extragalactic dipole identified in this work are a consequence of the heavy-mass composition assumption. 
Achieving such high amplitudes through the source distribution alone is challenging, unless only a small number of sources contributes significantly to the flux above $8$~EeV.

As suggested by other studies, a physically motivated source distribution following the large-scale structure of the Universe can also lead to a dipole anisotropy on Earth compatible with the measurements of the Pierre Auger Observatory~\citep{Harari2015, Ding21, Bister:2024ocm}. 
However, the mass composition of the cosmic-ray flux assumed in these studies is lighter than in the heavy-metal scenario. 


    \subsection{Arrival Directions at the Highest Energies}
  \label{sec:ArrivalDirectionsAtHE}

We backtrack the 100 most energetic events (above 78\,\EeV) measured by the Pierre Auger Observatory from~\citet{AugerMostEnergeticEvents} to the edge of the Galaxy, assuming all the particles are iron nuclei.
Similarly to Section~\ref{sec:Dipole}, we combine the coherent UF23 models with the turbulent field from the JF12 model, including the Planck Collaboration corrections and use multiple realizations of the turbulent field for three different coherence lengths. 
For each event, we backtrack 200 anti-iron nuclei, with their arrival directions smeared by a $1^{\circ}$ directional uncertainty, using a flat distribution around the reconstructed direction. 
The energy is smeared by a Gaussian distribution, accounting for the uncertainty in the reconstructed energy. 
This backtracking procedure is then repeated for an isotropic distribution of arrival directions, using the same energies as those of the detected events and corrected for the directional exposure of the Auger surface detector \citep{SOMMERS_SDexposure}.

The normalized histogram of the backtracked directions of the most energetic Auger events at the edge of the Galaxy is shown in the middle panel of Fig.~\ref{fig:ExtragalacticDipoleUF23}, in the Galactic coordinates. 
This plot combines the simulations of all the different realizations of the GMF. 
Assuming only iron nuclei, the sources of these particles are predominantly located in the direction of the Galactic anti-center. 
This behavior is strongly influenced by the low rigidity of the particles in the heavy-metal scenario and the GMF, which causes the de-magnification of sources in the direction of the Galactic Center.
A similar behavior is expected for an isotropic distribution of the arrival directions (see \change{the bottom panel of Fig.~\ref{fig:ExtragalacticDipoleUF23}})

The ratio of the distributions obtained from the backtracked Auger events and the backtracked isotropically distributed arrival directions on Earth is shown in the \change{Supplementary material~\ref{app:Dipole}}
.
The excess regions seem to follow the supergalactic plane, with the largest excess in the vicinity of the Centaurus A/M83 Group. 
However, these excesses may result from the low statistics of the observed extremely energetic events or from specific features of the UF23 model of the GMF. 
A more thorough investigation of these effects is beyond the scope of this paper.

    \section{Discussion}
  \label{sec:Discussion}

\subsection{Consistency with Other Studies}
  \label{sec:Consistency}

The interpretation of the data of the Pierre Auger Observatory becomes surprisingly consistent when assuming a simple but also extreme scenario in which the mass composition of cosmic rays at the flux suppression is dominated by iron nuclei.
After applying \DeltaXmax, the resulting moments of the logarithmic nuclear mass, $\ln A$, fall within the physical range expected for protons and He, N, Fe nuclei, see Fig.~\ref{fig:Umbrella}.
Furthermore, the value of $\sigma^{2}(\ln A)$ in the energy range $10^{18.5}\,\eV$ to $10^{19}\,\eV$ is consistent with the model-independent constraints from~\cite{MixedAnkle}.
For the \qgsii model, clear indications of too-shallow \Xmax predictions had already been claimed in the latter study, when the inferred primary-mass mixing was compared to that obtained from the fits of \Xmax distributions.
Moreover, in ~\cite{PierreAuger:2016use}, shifting the $\Xmax$ scale of models to some extent resulted in a more coherent picture of the mass-composition data from the Pierre Auger Observatory.

Interestingly, the mass estimates obtained from the surface-detector reconstructed $\Xmax$~\citep{DeltaMethod,PierreAuger:2024flk}, without calibration to the values measured by the fluorescence telescopes, are consistent with the mass-composition scenario proposed in this work.
However, this could be just coincidental, as the results of the two methods might be affected by the lack of muons in the simulated ground signal.

The interpretation of arrival directions measured by the Telescope Array experiment was found to be in better agreement with a heavier mass composition \citep{TA_HeavyArrivalDirections} than that obtained by the Pierre Auger Collaboration in \citet{PierreAuger:2017pzq}, which is again consistent with our heavy-metal scenario.
The modeled descriptions of the Auger dipole in arrival directions above $8\,\EeV$, with a source distribution following the large-scale structure \citep{AugerDipole24}, indicate that a heavier mass composition than that inferred using the standard model predictions could better explain the observed dipole amplitude, which is smaller than initially anticipated.

An indication of alleviation of the muon problem is also apparent when estimating the number of muons from the data of the Pierre Auger Observatory using a four-component shower universality model~\citep{Stadelmaier:2024pae}, applied to the combined fluorescence and surface detector data~\citep{PierreAuger:2023vcr}.
In the latter work, the number of hadronic shower particles at the ground is estimated from the independent fluorescence detector measurements of the shower development and the primary energy, thereby suppressing the sensitivity of the muon-scale estimation to the predicted \Xmax scale.

In the future, an update of the \eposlhc model of hadronic interactions will shift the $\Xmax$ scale for expectations deeper for the four primary particles, implying a heavier mass composition than previously assumed~\citep{Pierog:2023ahq}, approximately at the level of the \Xmax scale suggested in this work.

\subsection{Iron Nuclei in Cosmic-ray Source Candidates}\label{sec:Sources}

The \change{energy flux $J$} of cosmic rays above $10^{19.6}$~eV can be determined by integrating the cosmic-ray energy spectrum from \citet{SDEnergySpectrum2020}

\begin{equation} 
J(> 10^{19.6}{\,\eV}) = \int_{10^{19.6}{\,\eV}}^{10^{20.15}{\,\eV}} E\,J(E)\,dE. \end{equation}

\noindent We find that $J(> 10^{19.6}{\,\eV}) \sim 4.5 \times 10^{17}\,$\eV~km$^{-2}$~yr$^{-1}$~sr$^{-1}$. 
To calculate the luminosity density \change{$Q$} of these cosmic rays, we use

\begin{equation} Q(> 10^{19.6}{\,\eV}) \approx \frac{4\uppi\,J(> 10^{19.6}{\,\eV})}{D_{\rm loss}}, \end{equation}

\noindent where $D_{\rm loss}$ represents the characteristic energy loss length of the particles. 
Previous studies have discussed the propagation of iron nuclei with energies in the range $10^{19}\,\eV-10^{20}\,\eV$, finding $D_{\rm loss}\approx 100$~Mpc \citep{Allard2005,Jiang2021}. However, as we want the particles to arrive at the Earth retaining their iron identity, we adopt as $D_{\rm loss}$ the attenuation length used in our simulations above 40$\,\EeV$, which is approximately $30$~Mpc. 
With this, we find the required luminosity density to be $Q(> 10^{19.6}{\,\eV}) \approx 3 \times 10^{44}\,$erg~Mpc$^{-3}$~yr$^{-1}$.

It is reasonable to consider that, at the source, there may also be lighter elements that are accelerated up to lower energies due to their lower atomic numbers. 
Indeed, our results in the bottom panels of Fig.~\ref{fig:mass_spectrum} suggest a common origin for the nitrogen component as well. 
Assuming a particle injection spectrum $Q(E)=Q_{0}\,E^{-\gamma}$ with $\gamma=2$ at the sources, covering energies from $10^{18}$~eV to $10^{20}$eV, the normalization constant $Q_{0}$ can be determined by requiring the distribution to match $Q(> 10^{19.6}{\,\eV})$

\begin{equation}
    Q(> 10^{19.6}{\,\eV}) = \int_{10^{19.6}{\,\eV}}^{10^{20.15}{\,\eV}} E \, Q(E) \, dE = Q_{0} \, \ln\left(\frac{10^{20.15}{\,\eV}}{10^{19.6}{\,\eV}}\right).
\end{equation}

\noindent Using this value, we can calculate the integrated luminosity density of the sources over the energy range $10^{18}$~eV to $10^{20.15}$~eV, yielding $Q_{\rm CR-source} \approx 10^{45}$erg~Mpc$^{-3}$~yr$^{-1}$. 
Considering that the typical efficiency of the cosmic-ray acceleration is $10\%$ or less, the sources would need to have a luminosity density of at least $10^{46}$erg~Mpc$^{-3}$~yr$^{-1}$. 
This criterion narrows down the list of potential candidate sources for ultra-high-energy cosmic rays to blazars, radiogalaxies, hard X-ray active galactic nuclei (AGNs), and accretion shocks in galaxy cluster mergers, as only these sources are likely to meet the required luminosity density~\cite[see, e.g.,][]{Murase2019}. 
Both galaxy clusters and hard X-ray AGNs show evidence of high abundance of iron nuclei. 
In galaxy clusters, concentrations of iron nuclei are estimated to reach higher levels in the central regions \citep{Liu2020}. 
Similarly, hard X-ray AGNs also exhibit significant iron content, with inferred values exceeding the solar abundance \citep[][and references therein]{Komossa2001}. Observations of specific radiogalaxies suggest that iron concentrations peak near their centers as well, reaching values close to solar metallicity \citep[see, e.g.,][]{Werner2006}.
Among these sources, hard X-ray AGNs are the most abundant in the nearby Universe, while blazars are the rarest. Taking into account the iron attenuation length of $30$~Mpc, the most favored sources are hard X-ray AGNs and radiogalaxies. If, instead of assuming a power-law particle injection with an index $\gamma=2$, a spectral index $\gamma=1$ is considered, then $Q_{\rm CR-source} \approx 4\times10^{44}$ erg Mpc$^{-3}$ yr$^{-1}$ is obtained. This harder particle distribution would also make starburst galaxies viable candidates, though only marginally, as they would just meet the minimum energy requirement.

To estimate the level of expected purity of a beam of iron nuclei at Earth, we consider uniformly distributed sources from 3~Mpc up to 150~Mpc with the source density $\rho=10^{-4}\rm{Mpc^{-3}}$ producing iron nuclei with an energy spectrum following the power-law with spectral index $\gamma=2$ and an exponential rigidity cut-off for two values of $R_{\rm{cut}}$ of 3~EV and 5~EV. 
Using the simulated energy losses on the cosmic microwave background and the extragalactic background light \citep{EBL_Gilmore}, we find that \avg{\ln A} on Earth above $40$~EeV is above $\sim3.8$ and $\sigma^{2}(\ln A)$ is below $\sim 0.03$. This shows that the propagation has a small effect on the mass composition above 40~EeV, preserving the scenario suggested in this article.



  \section{Conclusions}
We have presented a data-driven mass-composition scenario for ultra-high-energy cosmic rays, in which we attempt to achieve a physically consistent interpretation of the depth of the shower maximum (\Xmax) measured by the Pierre Auger Observatory.
To this end, we assume an extreme astrophysical benchmark scenario in which the cosmic-ray flux above $\approx40\,\EeV$ consists purely of iron nuclei and allows for shifts in the expected \Xmax scale of the two models of hadronic interactions, \qgsii and \sib{2.3d}.
The resulting shifts of the predicted \Xmax scale to deeper values closely match those obtained from joint fits of \Xmax and the ground signal distributions at $3\,\EeV-10\,\EeV$ \citep{PierreAuger:2024neu}.
Such a shift of the predicted \Xmax scale might be explained by recent improvements in air-shower modelling \citep{Pierog:2023ahq}. 
Consequently, the changes in the mean and variance of $\ln A$ shift the measured data at $3\,\EeV-100\,\EeV$ within the region of expected combinations of protons and He, N, and Fe nuclei, contrary to the standard interpretation using unmodified model predictions. 
The variance of $\ln A$ is then consistent with the model-independent constraints on the broadness of the cosmic-ray mass composition at energies $3\,\EeV-10\,\EeV$ \citep{MixedAnkle}.
Furthermore, we discussed the implications of this scenario on the consistency of the decomposed energy spectrum, hadronic interaction studies, and backtracked arrival directions.

In the presented heavy-metal scenario, the flux suppression of cosmic rays is consistent with a rigidity cut-off approximately at $2\,\text{EV}$.
Consequently, nitrogen and iron nuclei starting to disappear from the cosmic rays at the same rigidity could explain the \emph{instep} feature of the cosmic-ray energy spectrum.
The deficit of muons predicted by \qgsii and \sib{2.3d}, compared to direct measurements of the muon signal, is alleviated from $\sim30\%-50\%$ to $\sim20\%-25\%$, approximately independently of the primary energy.
There is an indication that the inelastic p-p cross-section or elasticity needs to be modified in \sib{2.3d} or \qgsii within the heavy-metal scenario at $10^{18.0-18.5}\,\eV$, however, the overall description of the \Xmax distribution remains reasonable within the statistical uncertainties.
Considering the observed dipole anisotropy of ultra-high-energy cosmic rays at energies above $8\,\EeV$, we confirm that this observation is consistent with a possible extragalactic dipolar distribution of cosmic-ray sources within the heavy-metal scenario at the $2\sigma$ level and even at the $1\sigma$ level for very high extragalactic amplitudes (above $\sim 40\%$).
Assuming only iron nuclei, the arrival directions of the most energetic Auger events, when backtracked through the Galactic magnetic field, point towards the Galactic anticenter region, which is consistent with the expectations from isotropic arrival directions at the Earth.
The estimated integrated luminosity density of the sources within the heavy-metal scenario suggests that only very powerful objects, such as hard X-ray AGNs, could explain the origin of the ultra-high-energy cosmic rays.

\begin{acknowledgments}
The work was supported by the Czech Academy of Sciences: LQ100102401, Czech Science Foundation: 21-02226M, Ministry of Education, Youth and Sports, Czech Republic: FORTE CZ.02.01.01/00/22\_008/0004632, German Academic Exchange service (DAAD PRIME). 
The authors are very grateful to the Pierre Auger Collaboration for discussions about this work, especially to G.~R.~Farrar and M.~Unger.

\end{acknowledgments}

\bibliography{bibtex}{}
\bibliographystyle{aasjournalv7}

\appendix

\section{Purity of primary beam}
\label{app:PurityTests}
\begin{figure*}
    \includegraphics[width=0.48\textwidth]{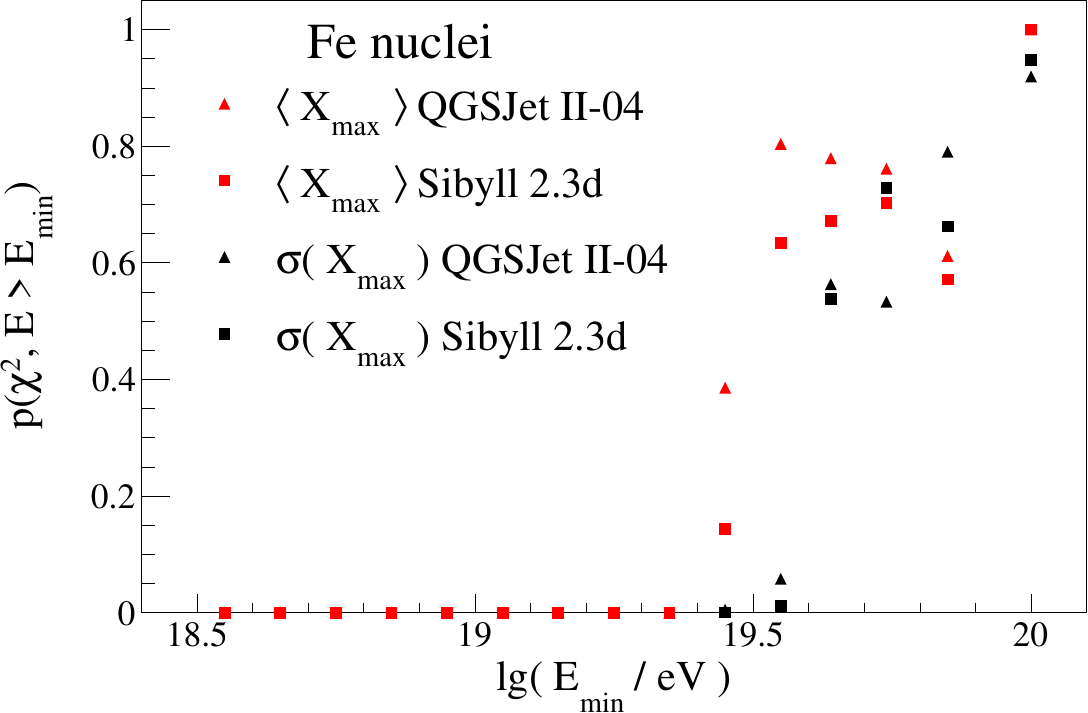}
    \hfill
    \includegraphics[width=0.48\textwidth]{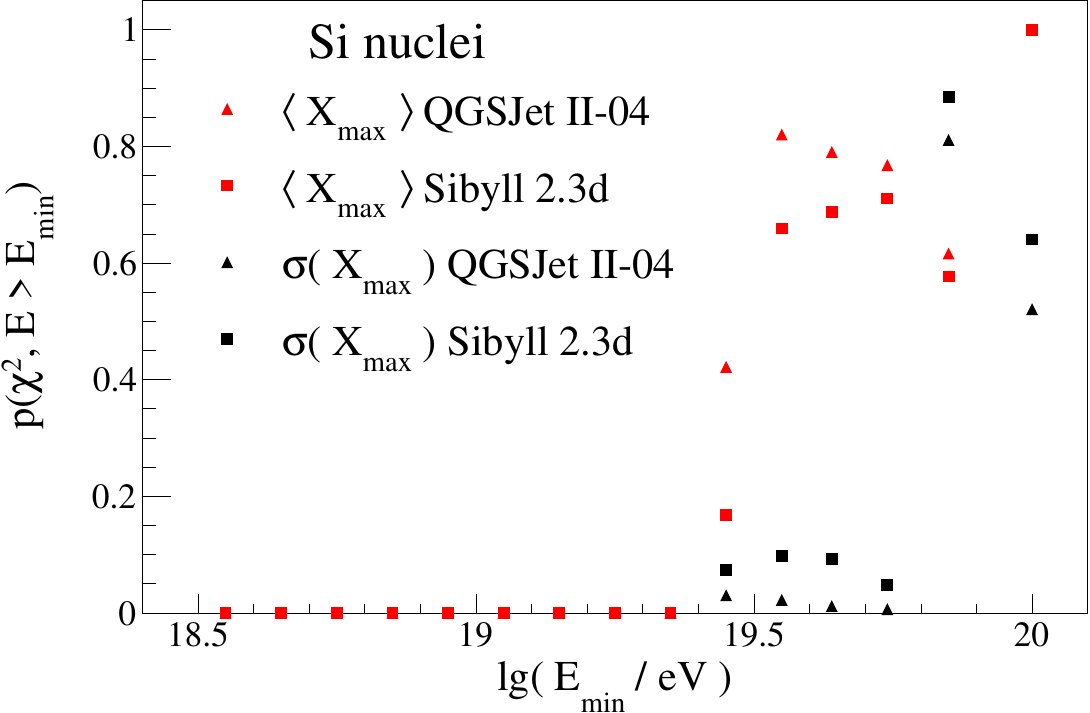}
    \caption{Test of beam purity for iron nuclei (left) and silicon nuclei (right) above the energy $E_\text{min}$ using the Auger DNN data \citep{PierreAuger:2024flk}. The $p$-values of the $\chi^{2}$ test for a constant elongation rate of a single primary type above $E_\text{min}$ (in red) and for \Xmax fluctuations consistent with pure nuclei above $E_\text{min}$ (in black).}
    \label{fig:BeamPurity}
  \end{figure*}

We apply the $\chi^{2}$ test to check the consistency of a pure beam of Fe and Si nuclei above a given energy, as shown in Fig.~\ref{fig:BeamPurity}. 
For this, we use the elongation rate and \Xmax fluctuations predicted by the \qgsii and \sib{2.3d} models, comparing them with the Auger DNN data \citep{PierreAuger:2024flk}.
In case of the elongation-rate test, the energy evolution was fitted with the model prediction, allowing freedom in the \Xmax scale.
These tests show compatibility with a pure beam of Fe nuclei above $10^{19.6}\,\eV$ in both \Xmax moments.
In case of the Si nuclei, a consistent description of both \Xmax moments can only be achieved at higher energies than for Fe nuclei, where the event statistics become very low.
This inconsistency is caused by the poor description of \Xmax fluctuations, which would be even worse for lighter nuclei than Si.
Note also that in the case of pure Si nuclei at the highest energies, the obtained \Xmax shifts are $(36\pm1)~\gcm$ and $(12\pm1)~\gcm$ for \qgsii and \sib{2.3d}, respectively, which is in tension with the results from \citep{PierreAuger:2024neu}, even when accounting for the systematic uncertainties, which are highly correlated between these two results.

\section{Functional forms of the energy evolution of primary fractions}
\label{app:functions_fractions}
We use two different functional forms to describe the energy evolution of the fitted primary fractions shown in the top panels of Fig.~\ref{fig:mass_spectrum}.
The Gauss $\times$ exponential function provides a purely empirical description of the energy evolution of primary fractions, while the power-law function with a simple exponential cutoff allows for a more physically motivated description of the energy evolution of primary fractions, as discussed in \citet{PierreAuger:2017pzq}.
\subsection{Gauss $\times$ exponential function}
\label{sec:GaussExponentialForm}
The smoothed description of the energy evolution of the primary fractions i = p, He, N, Fe is obtained by a log-likelihood minimization using the following functions
\begin{equation}
    \hat{f_\text{i}}=A_\text{i}\exp\left[-\frac{(\lg( E/\eV)-\mu_\text{i})^2}{2\sigma_\text{i}^2}\right] \times f_\text{exp},   
\end{equation}
\noindent where
\begin{equation}
    f_\text{exp} = \frac{1}{1+\exp\left[10\,(\lg( E/\eV)-W_\text{i})\right]},
\end{equation}
\noindent if i = p, He or N. For Fe nuclei, we set $f_\text{exp}=1$.
The final primary functions, which are minimized simultaneously, are normalized to $\sum f_{i}=1$ using
\begin{equation}
    f_\text{i} = \hat{f_\text{i}} / \sum \hat{f_\text{i}}.
\end{equation}
The resulting fitted parameters are given in Tab.~\ref{TabGaussExp}.

\begin{table*}
  \renewcommand{\arraystretch}{1.2}
      \caption{Parameters of the Gauss $\times$ exponential function fitted to the energy evolution of the primary fractions in the top panels of Fig.~\ref{fig:mass_spectrum} for \qgsii (left) and \sib{2.3d} (right).}
      \begin{tabular}{l|cccc||cccc}
        \hline\hline
    & $A_{i}$ & $\mu_{i}$ & $\sigma_{i}$ & $W_{i}$ & $A_{i}$ & $\mu_{i}$ & $\sigma_{i}$ & $W_{i}$ \\
$i = \text{p}$ & 1.248 & 18.237 & 0.405 & 19.659 & 0.001 & 18.454 & 0.372 & 19.509 \\ 
$i = \text{He}$ & 1.077 & 18.486 & 0.624 &  19.659 & 0.004  & 1254.368 & 640.098 & 19.509 \\
$i = \text{N}$ & 1.128 & 18.805 & 0.288 & 19.659 & 0.003 & 18.874 & 0.405 & 19.509 \\
$i = \text{Fe}$ & 2.739 & 20.089 & 11.368 & -- & 25.661 & 35.606  & 3.95  & -- \\
        \hline\hline
      \end{tabular}
      \label{TabGaussExp}
  \end{table*}

\subsection{Power-law function with a simple exponential cutoff}
We also considered a description of the energy evolution of the primary fractions using a power-law function with a simple exponential cutoff above the $\lg$ energy $Y_{i}$, except for Fe nuclei.
For i = p, He and N nuclei, we use for the $\chi^{2}$ minimization of the primary fractions
\begin{equation}
    \lg f_{i}= a_{i}\, \lg( E/\eV) + b_{i} + \lg(e)\,( 1 - 10^{\lg( E/\eV)-Y_{i}} ).
\end{equation}
In case of the Fe nuclei, we use power-law functions with a break at $\lg$ energy $Y_\text{Fe}$ and force the fraction of iron nuclei to be 1 above the energy $10^{19.6}\eV$
\begin{align}
        \lg f_\text{Fe} &= a_\text{Fe}\,(\lg( E / \eV) ) - Y_\text{Fe}) + c_\text{Fe}\,(Y_\text{Fe}-19.6),\,\,\,\,\,\, &\text{if}\, \lg( E/\eV ) < Y_\text{Fe},\\
    \lg f_\text{Fe} &= a_\text{Fe}\,(\lg( E/\eV) - Y_\text{Fe}), \,\,\,\,\,\, &\text{if}\, Y_\text{Fe} \le \lg( E/\eV) \le 19.6, \\
    \lg f_\text{Fe} &= 0, \,\,\,\,\,\, &\text{if}\, 19.6 \le \lg( E/\eV).
\end{align}
The resulting fitted parameters are given in Tab.~\ref{TabLPowerLaws}.
Note that these parameterizations do not satisfy $\sum f_{i}=1$ within $\approx15\%$.

    \begin{table*}
  \renewcommand{\arraystretch}{1.2}
      \caption{Parameters of the power-law functions fitted to the energy evolution of primary fractions in the top panels of Fig.~\ref{fig:mass_spectrum} for \qgsii (left) and \sib{2.3d} (right).}
      \begin{tabular}{l|cccc||cccc}
        \hline\hline
    & $a_{i}$ & $b_{i}$ & $c_{i}$ & $Y_{i}$ & $a_{i}$ & $b_{i}$ & $c_{i}$ & $Y_{i}$ \\
$i = \text{p}$ & 0.104  & -2.87 & -- & 18.87  & 0.644  & -12.60 & -- & 18.48 \\
$i = \text{He}$ & 0.402 & -8.41 & -- & 19.03 & 0.144 & -3.58 & -- & 18.75  \\
$i = \text{N}$ & 3.479 & -65.30  &  -- & 18.37  & 1.022 & -19.67  &  -- & 18.91  \\
$i = \text{Fe}$ &  -0.214 & --  & 0.325  & 18.55  & -0.270 & --  & 0.467  & 18.54  \\
        \hline\hline
      \end{tabular}
      \label{TabLPowerLaws}
  \end{table*}

\section{\Xmax distributions of fractions fit}
\label{app:XmaxMassFits}
\begin{figure}[htp]
    \centering
    \begin{tabular}{cccc}
        \includegraphics[width=0.24\textwidth]{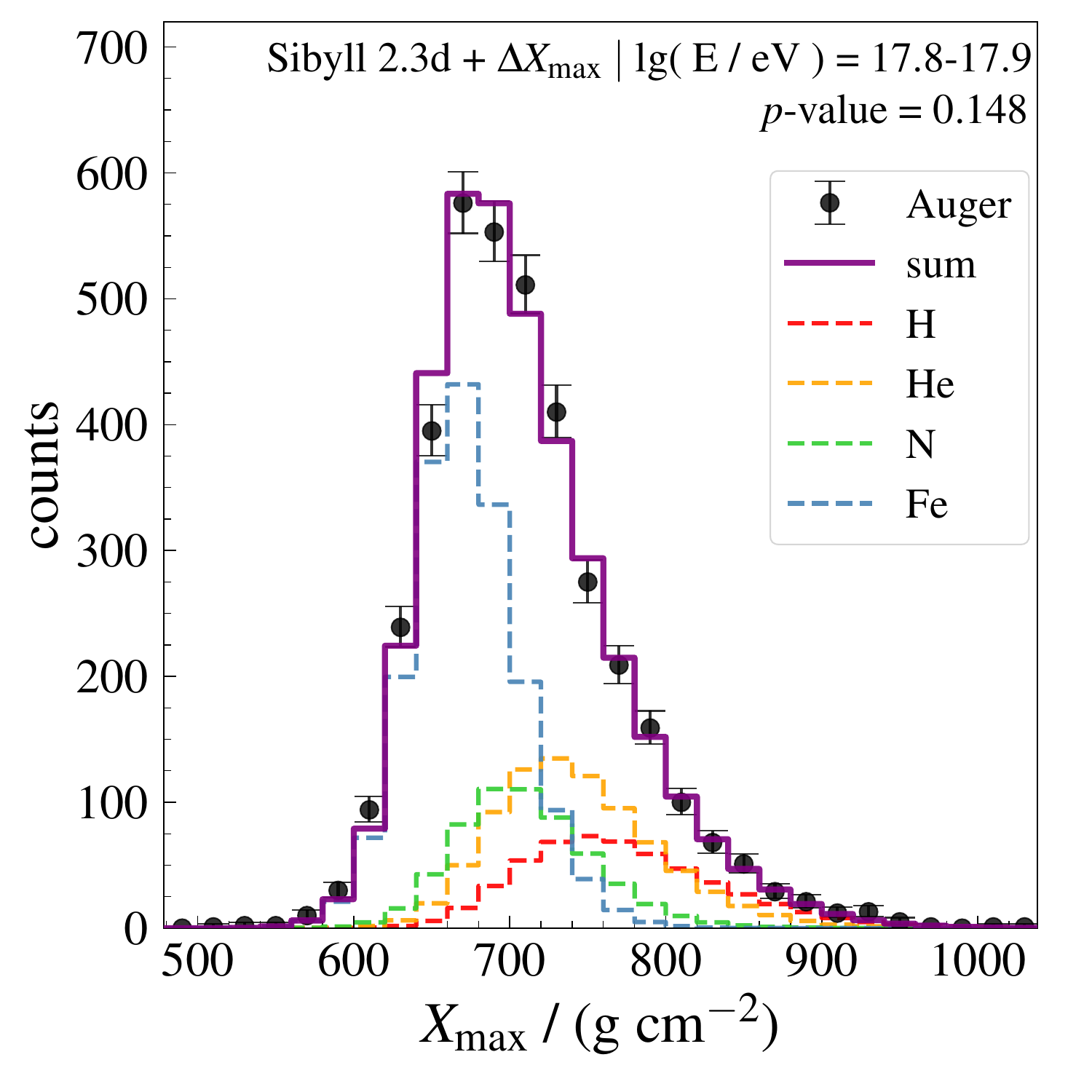} &
        \includegraphics[width=0.24\textwidth]{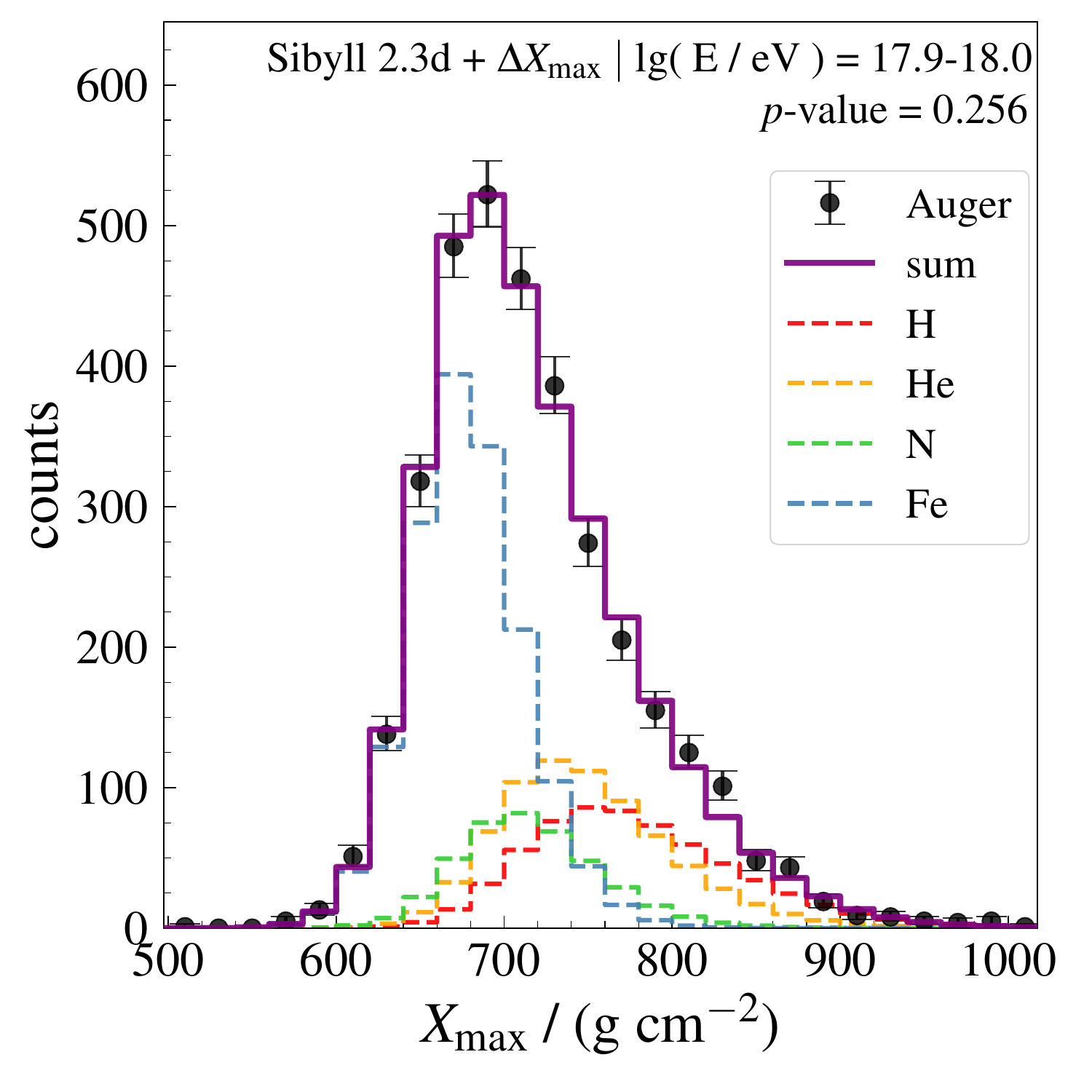} &
        \includegraphics[width=0.24\textwidth]{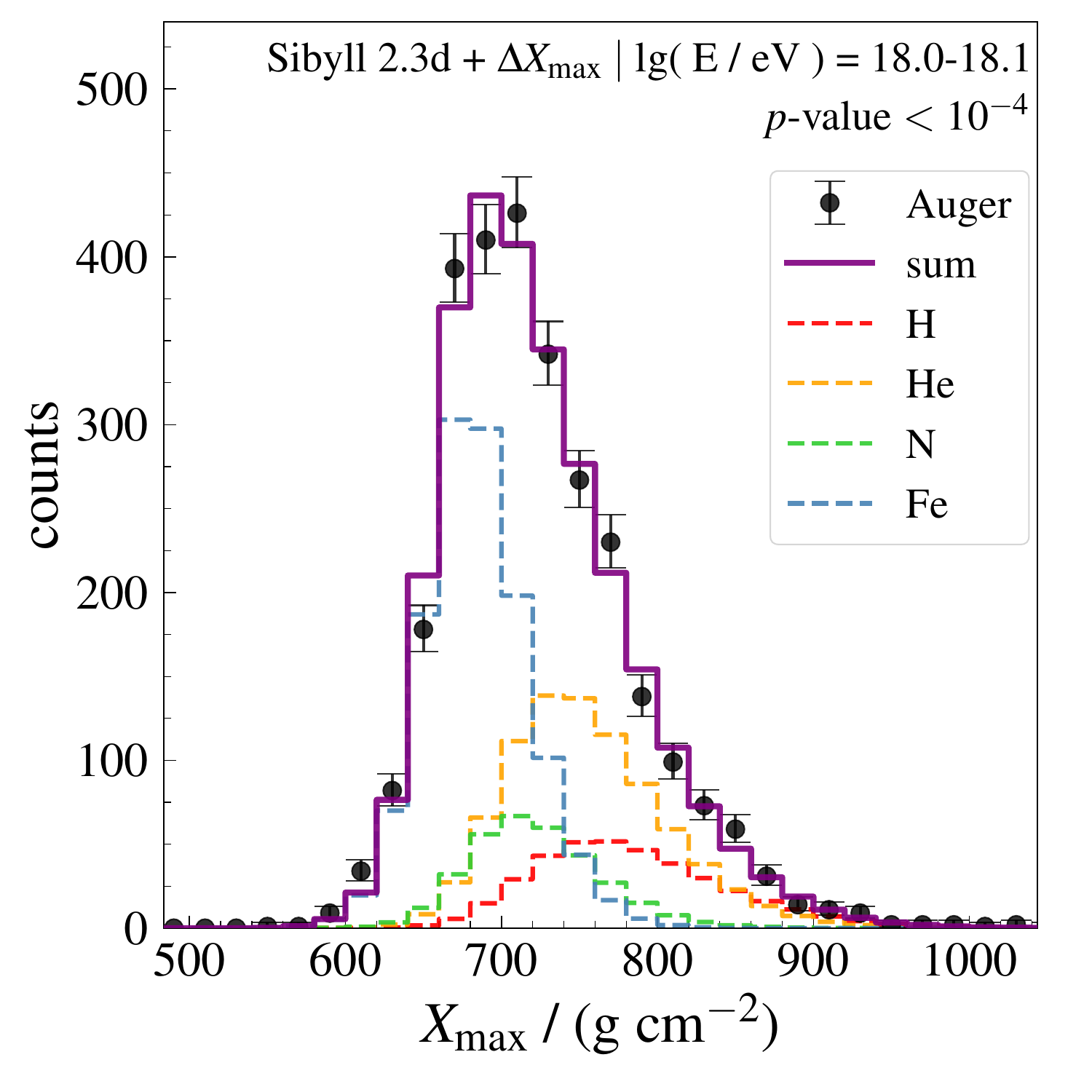} &
        \includegraphics[width=0.24\textwidth]{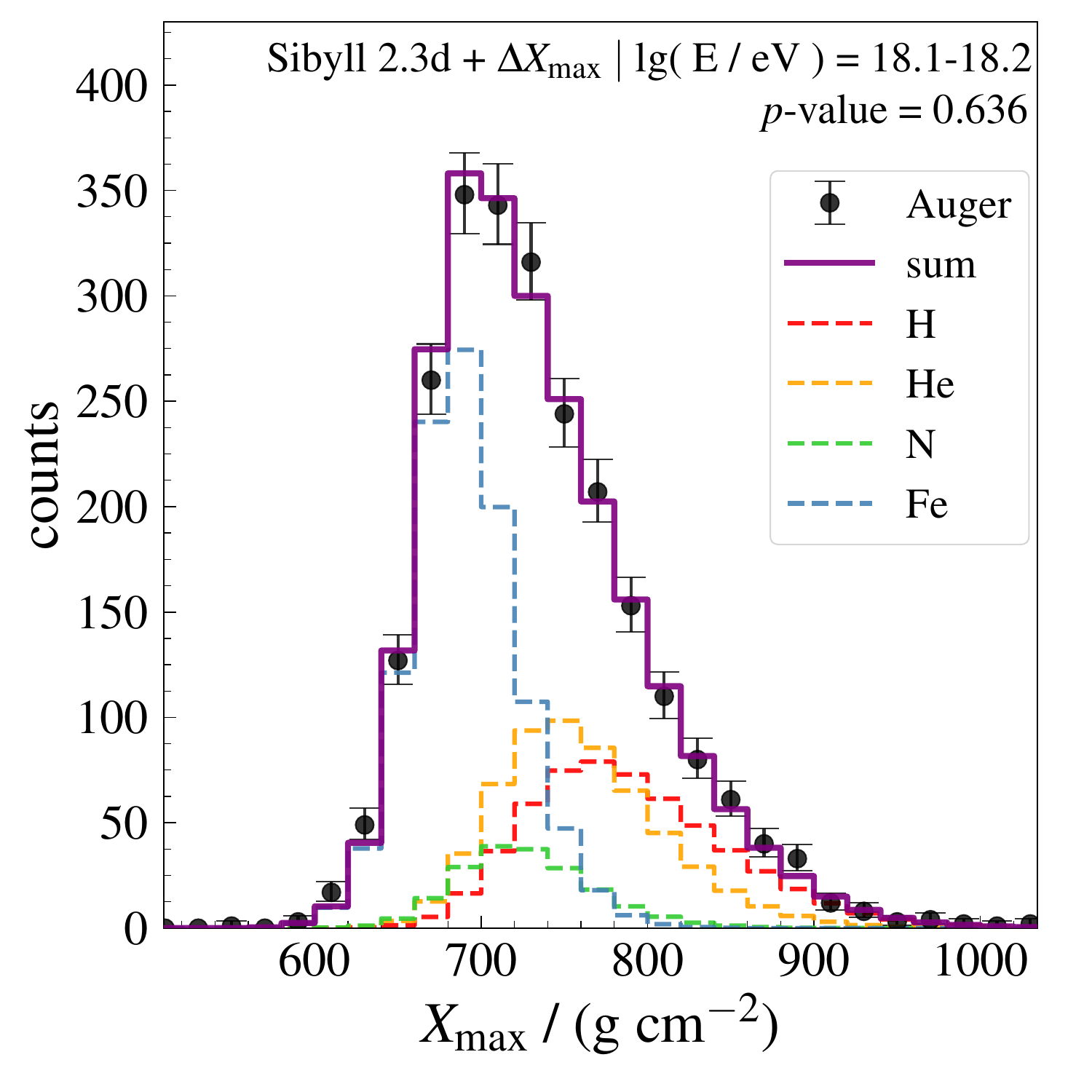} \\

        \includegraphics[width=0.24\textwidth]{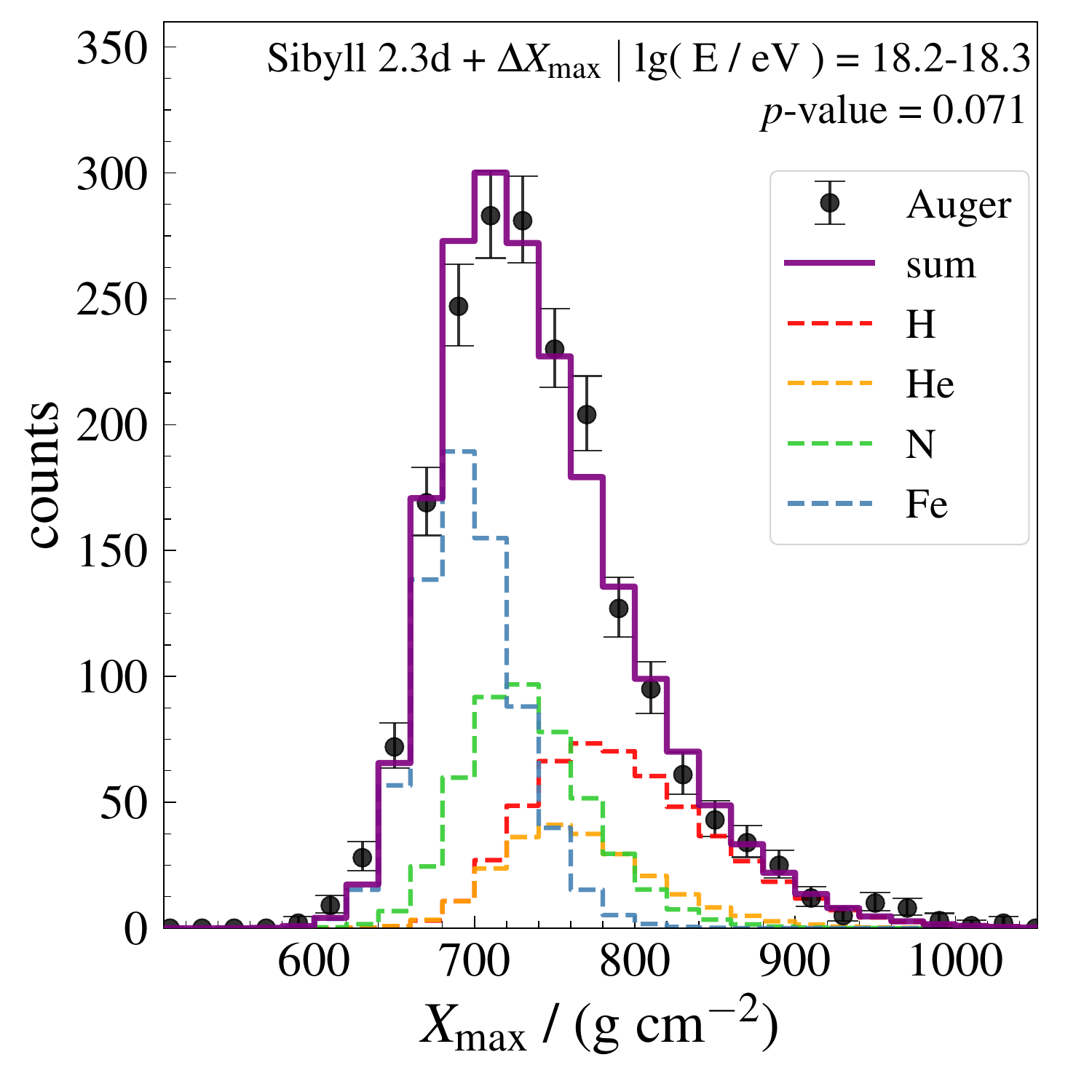} &
        \includegraphics[width=0.24\textwidth]{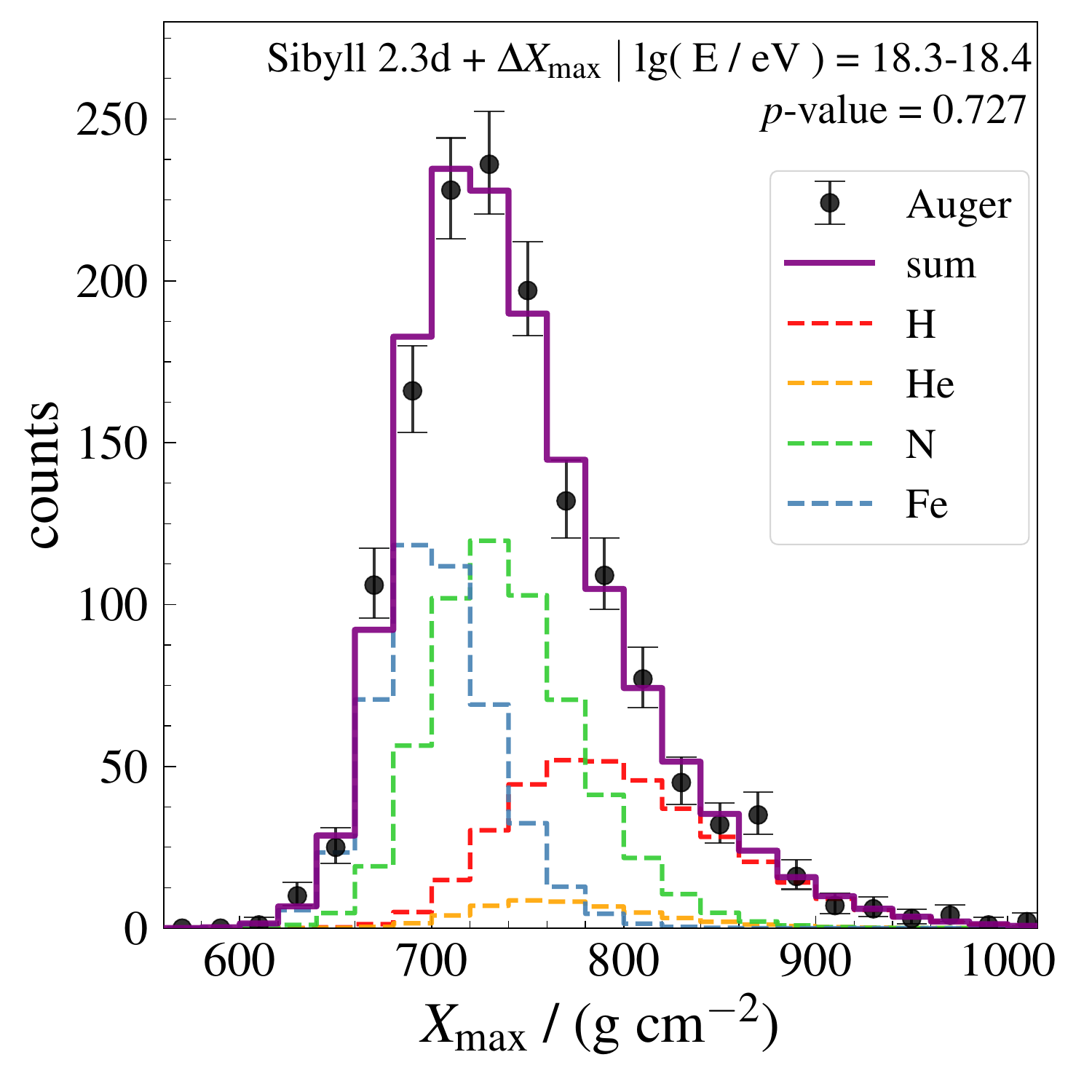} &
        \includegraphics[width=0.24\textwidth]{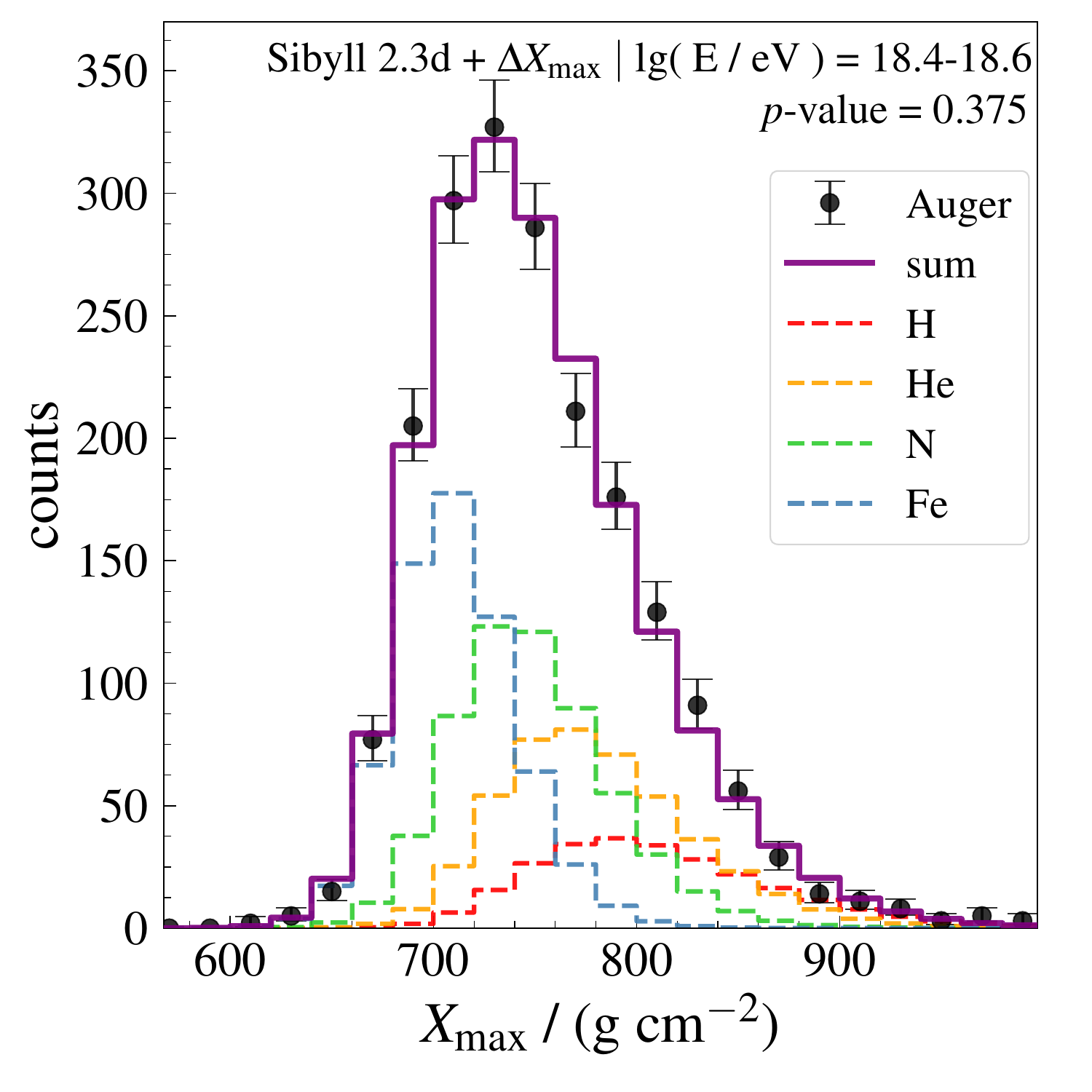} &
        \includegraphics[width=0.24\textwidth]{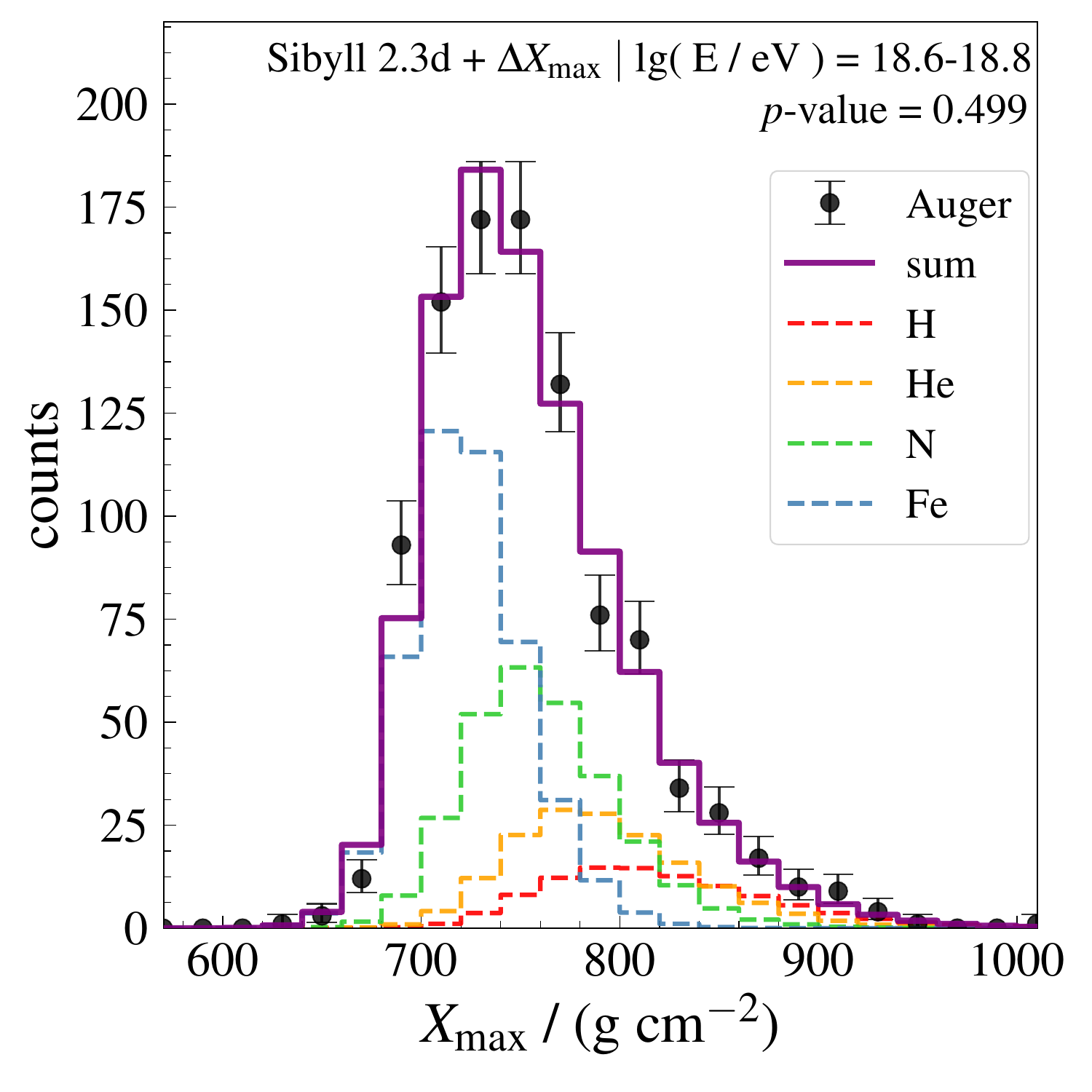} \\
        \includegraphics[width=0.24\textwidth]{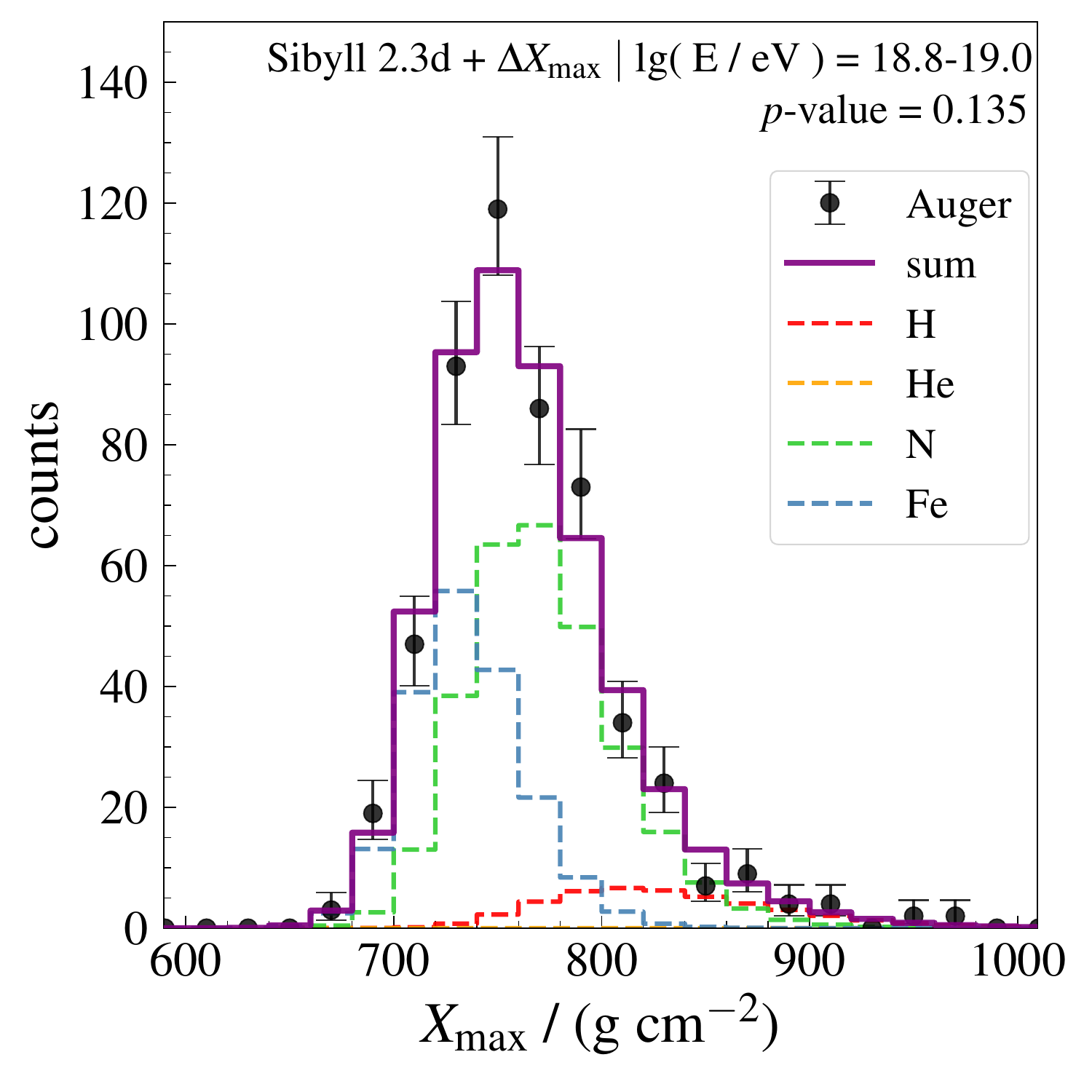} &
        \includegraphics[width=0.24\textwidth]{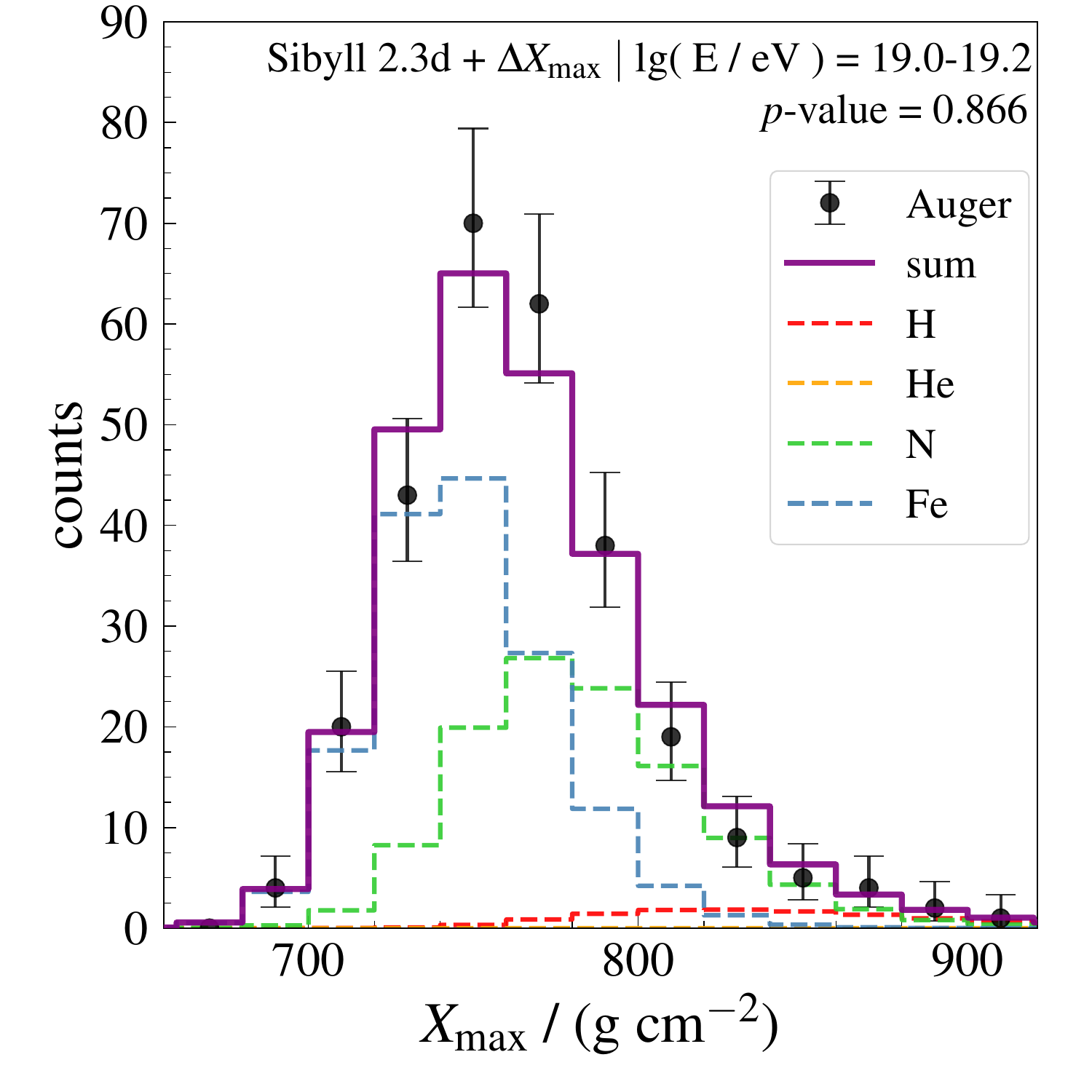} &
        \includegraphics[width=0.24\textwidth]{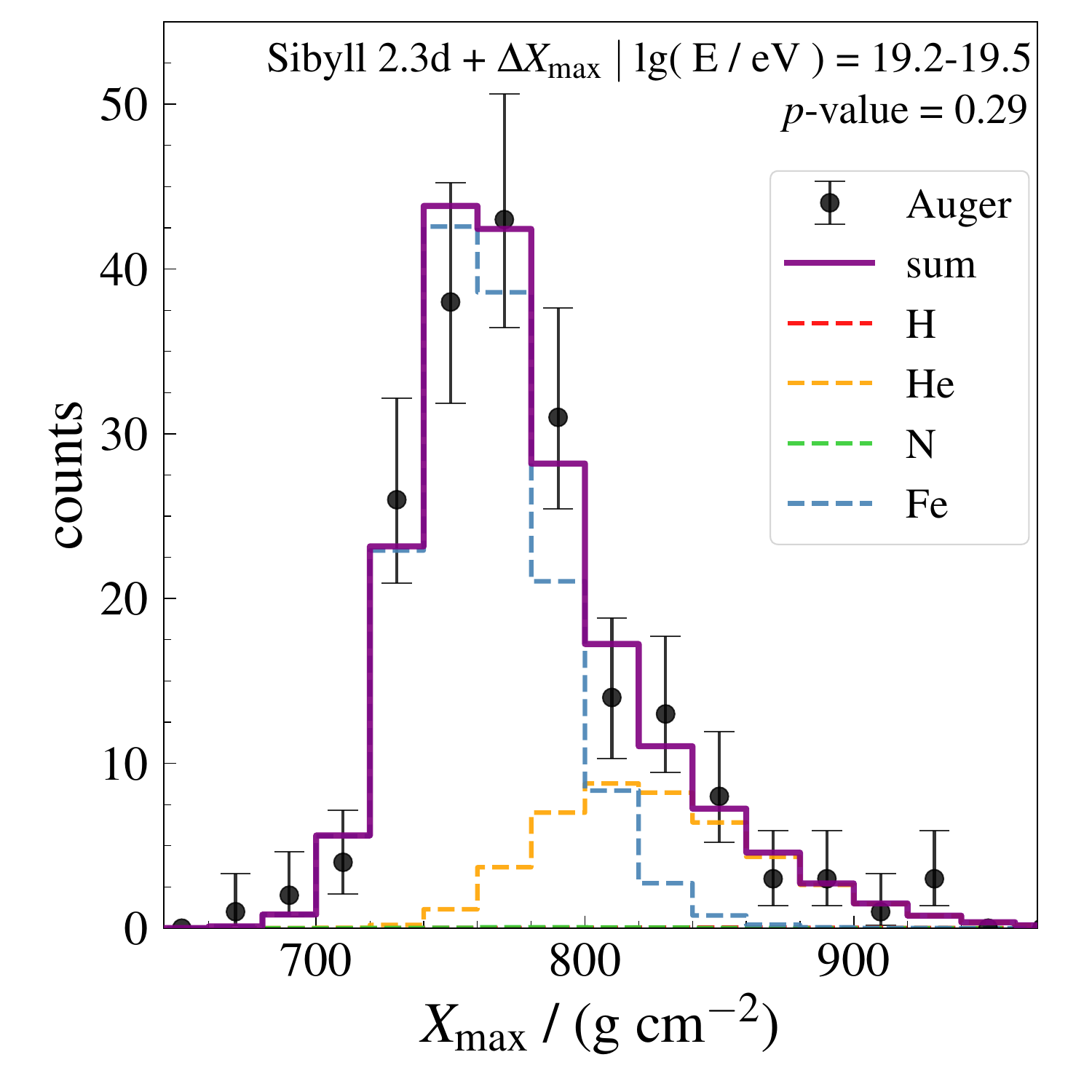} &
        \includegraphics[width=0.24\textwidth]{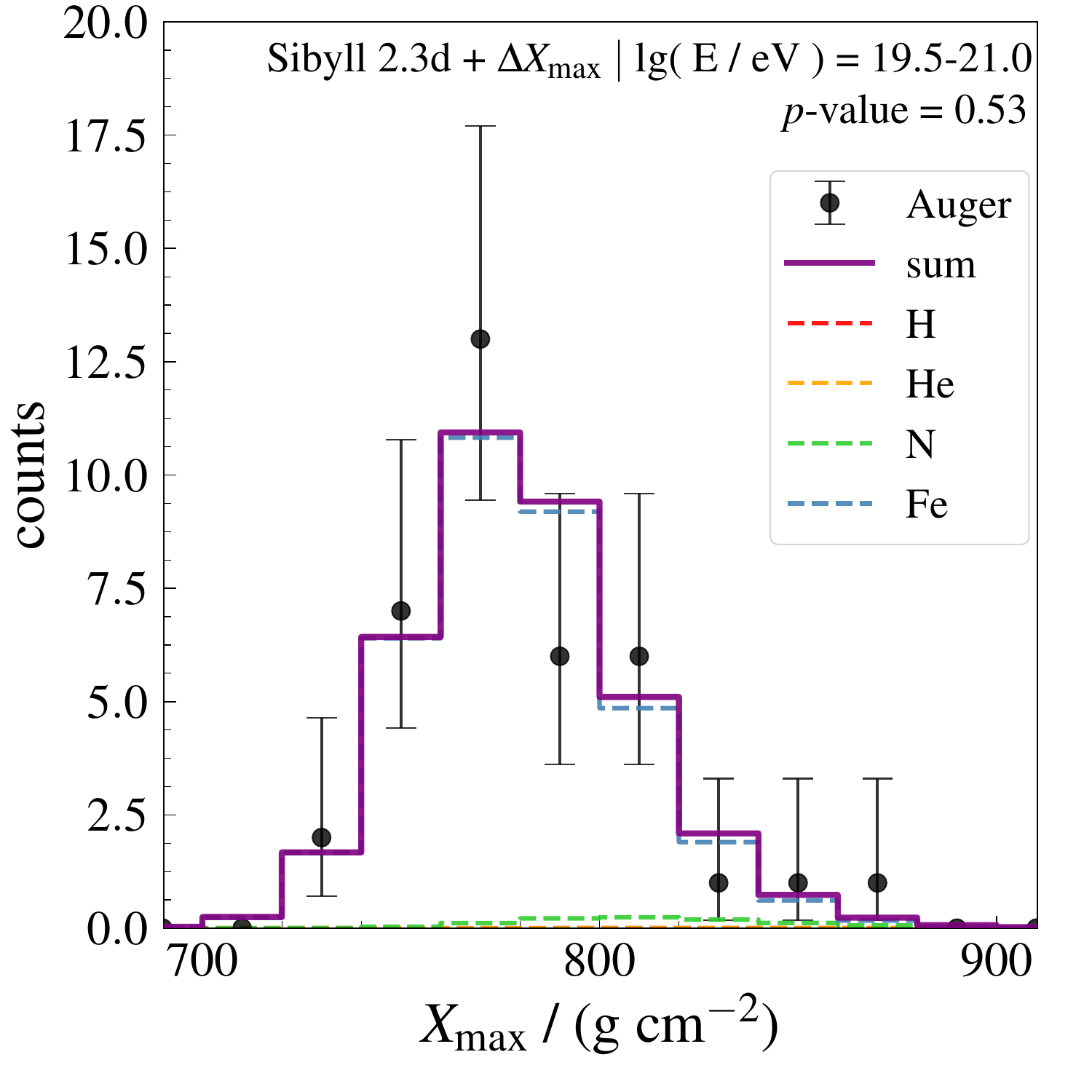} \\
    \end{tabular}
    \caption{Distributions of \Xmax obtained from fits in each energy bin using the model \sib{2.3d}$+\DeltaXmax$. 
    The plot shows the total fit (\textit{sum}) alongside the individual contributions of particle species. The $p$-value of the fit and the energy bin for which the fit is performed are indicated in each panel.}
    \label{fig:Xmax-distr-sibyll23d}
\end{figure}

\begin{figure}[htp]
    \centering
    \begin{tabular}{cccc}
        \includegraphics[width=0.24\textwidth]{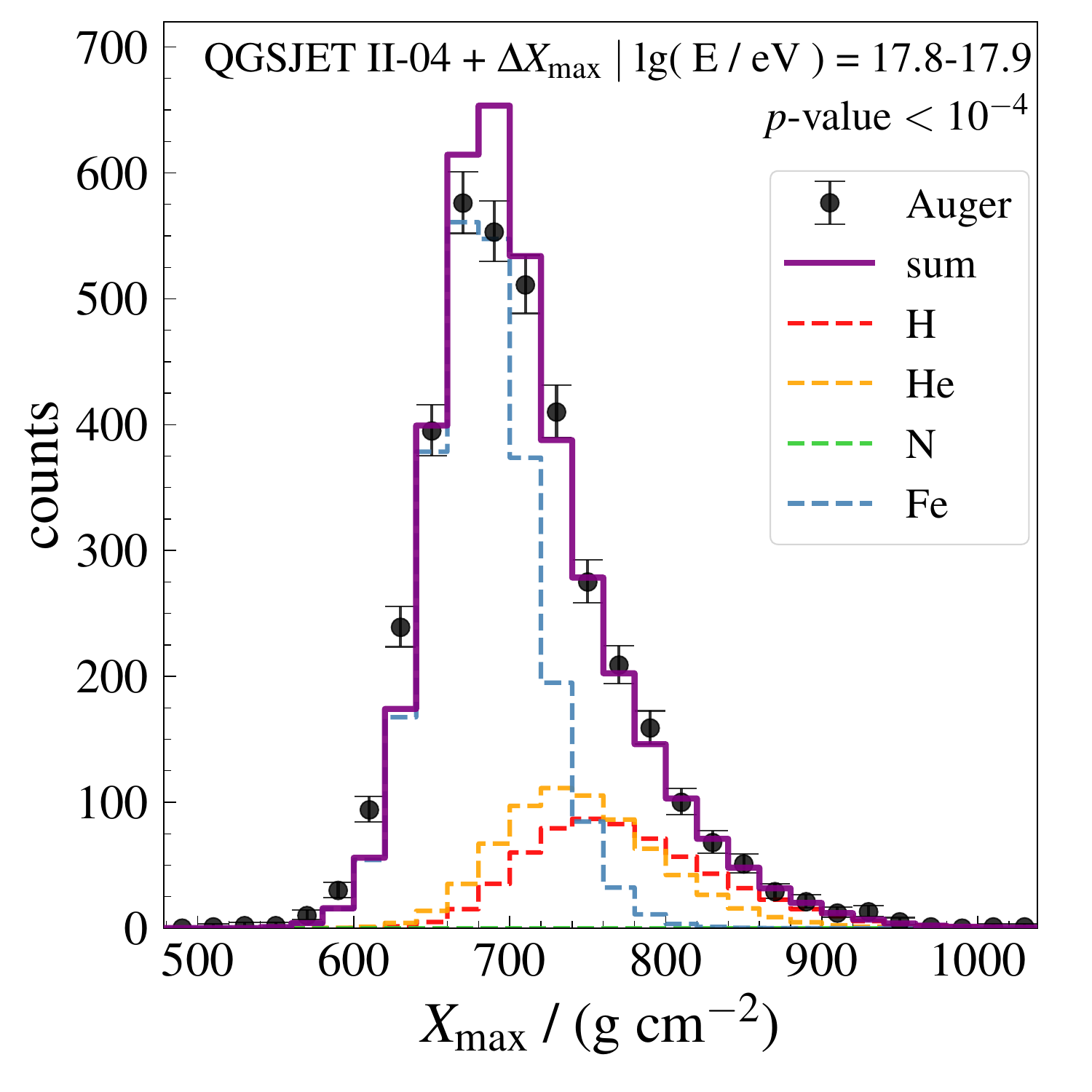} &
        \includegraphics[width=0.24\textwidth]{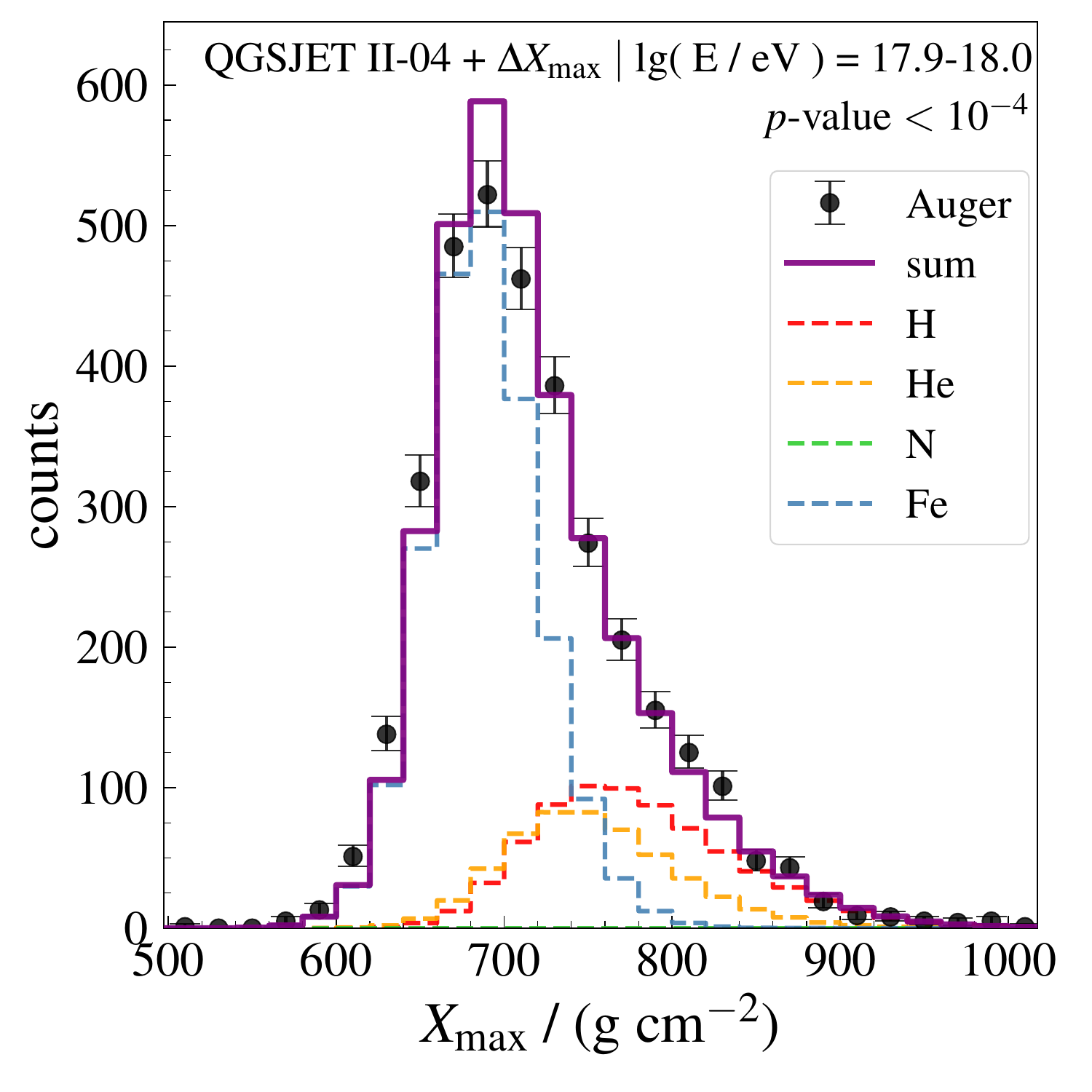} &
        \includegraphics[width=0.24\textwidth]{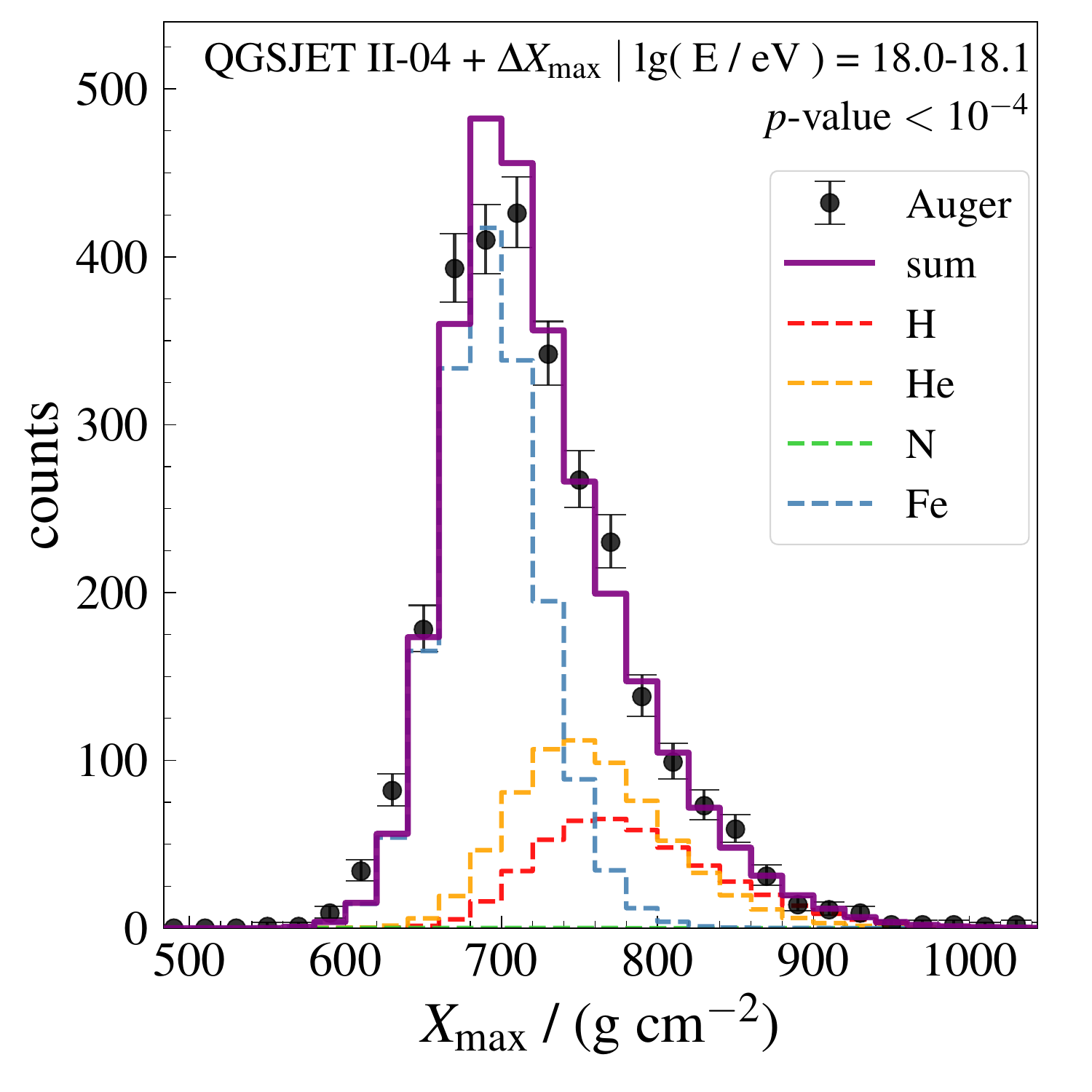} &
        \includegraphics[width=0.24\textwidth]{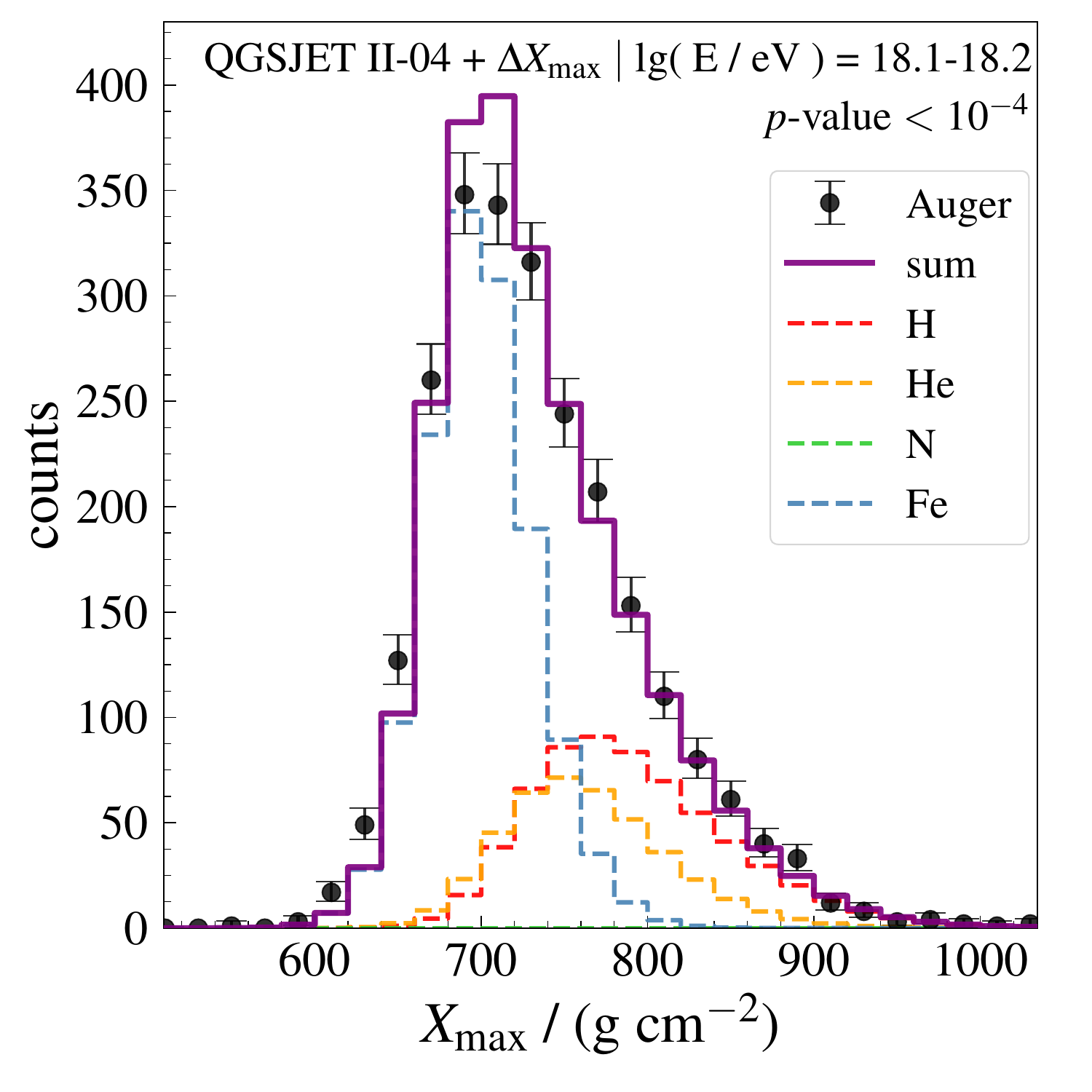} \\
        \includegraphics[width=0.24\textwidth]{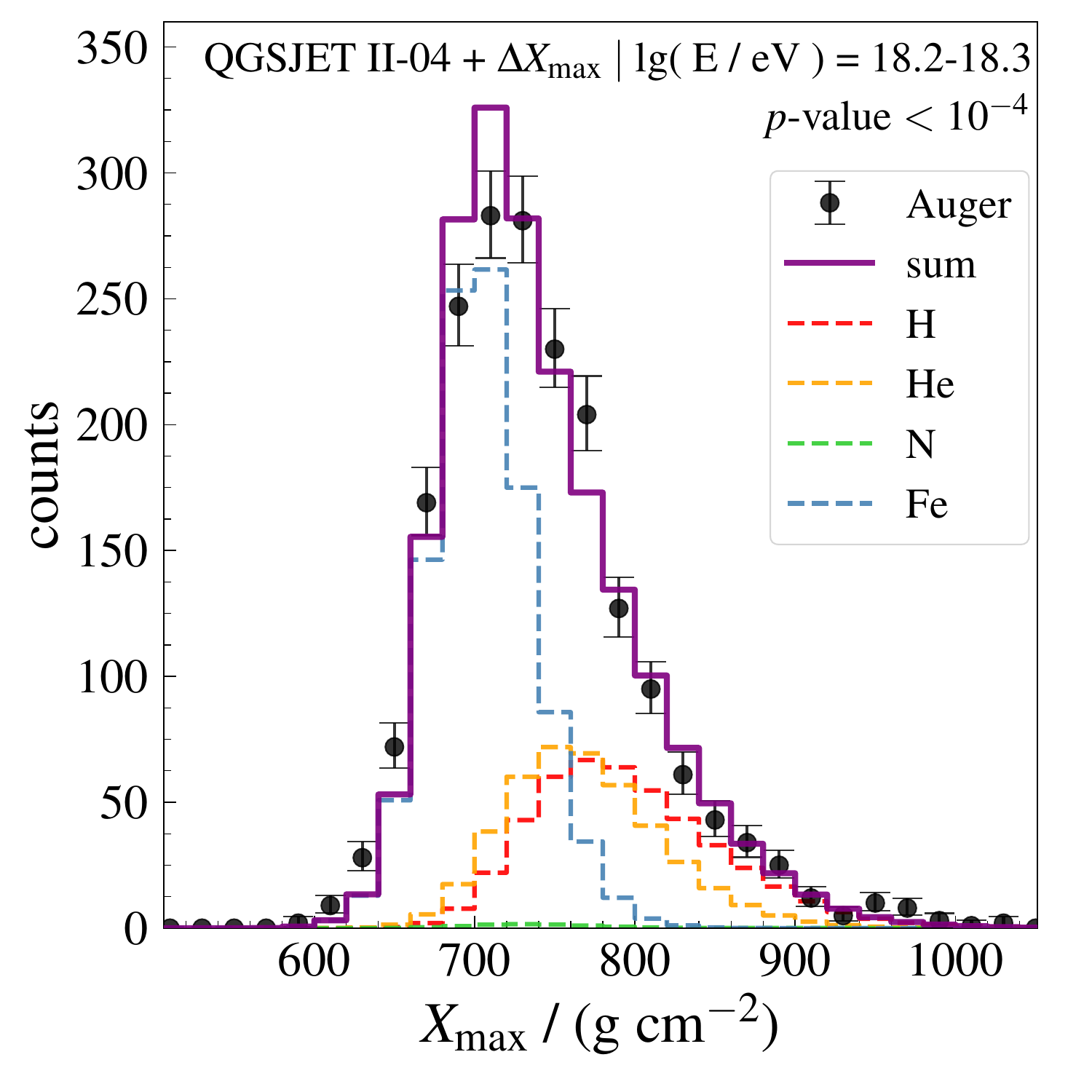} &
        \includegraphics[width=0.24\textwidth]{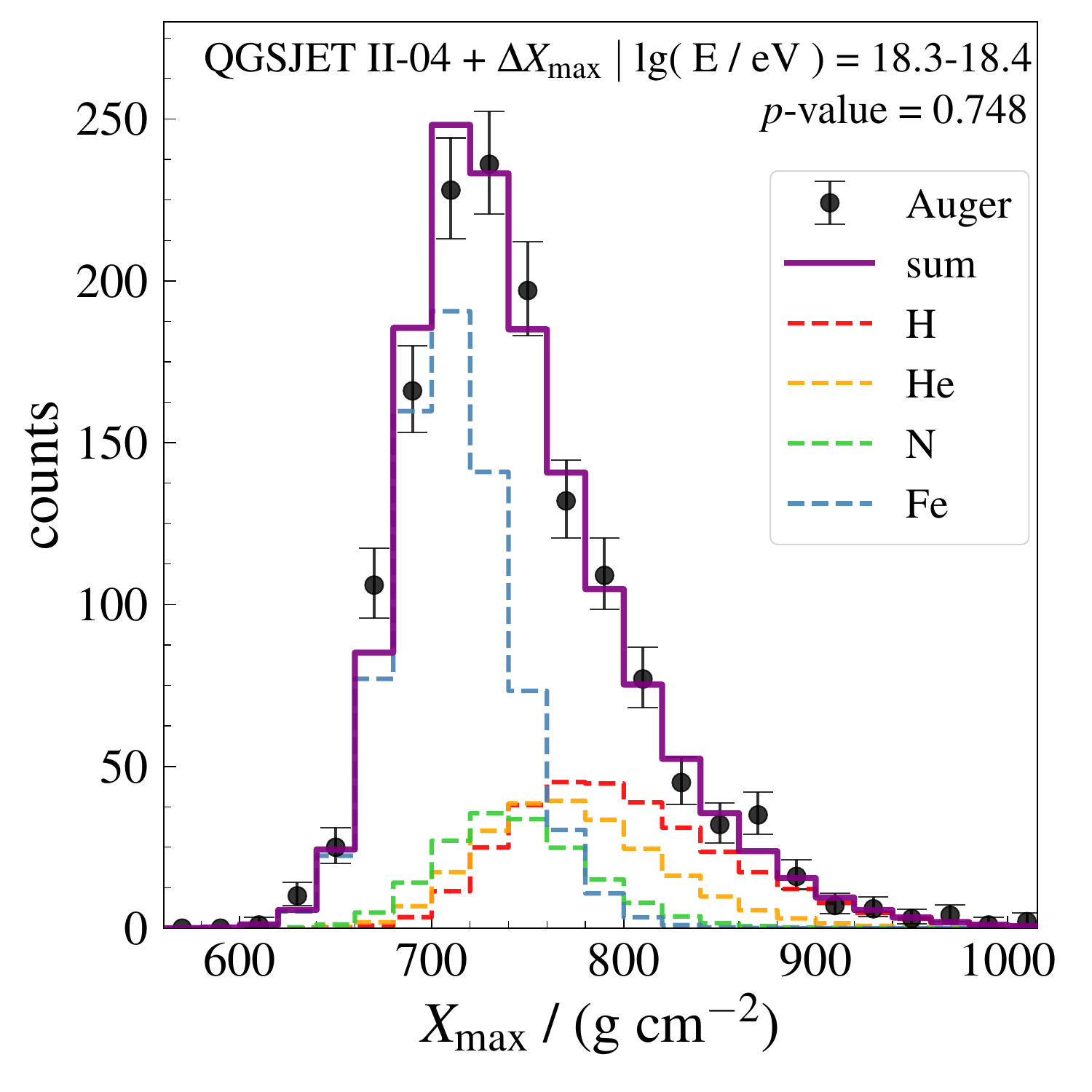} &
        \includegraphics[width=0.24\textwidth]{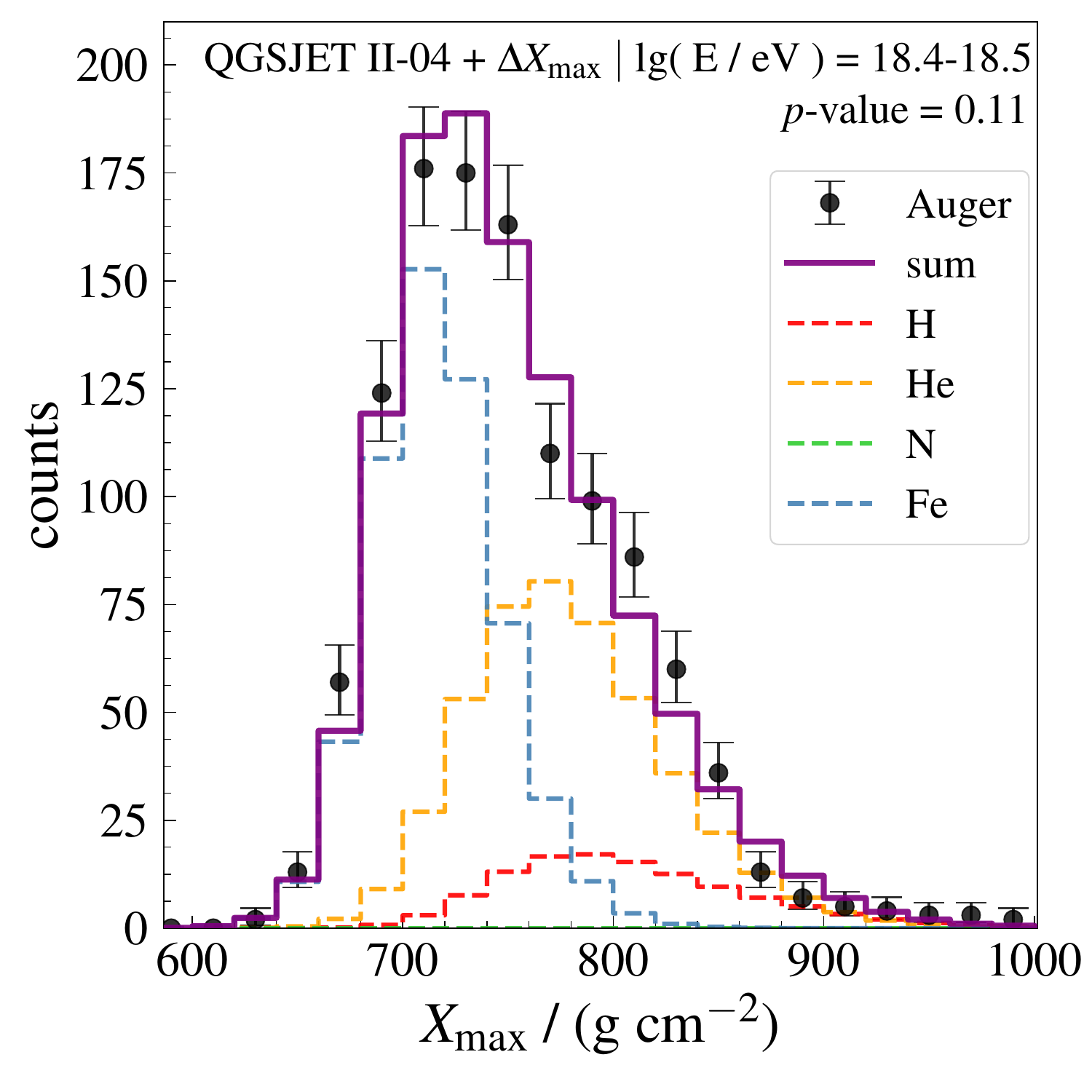} &
        \includegraphics[width=0.24\textwidth]{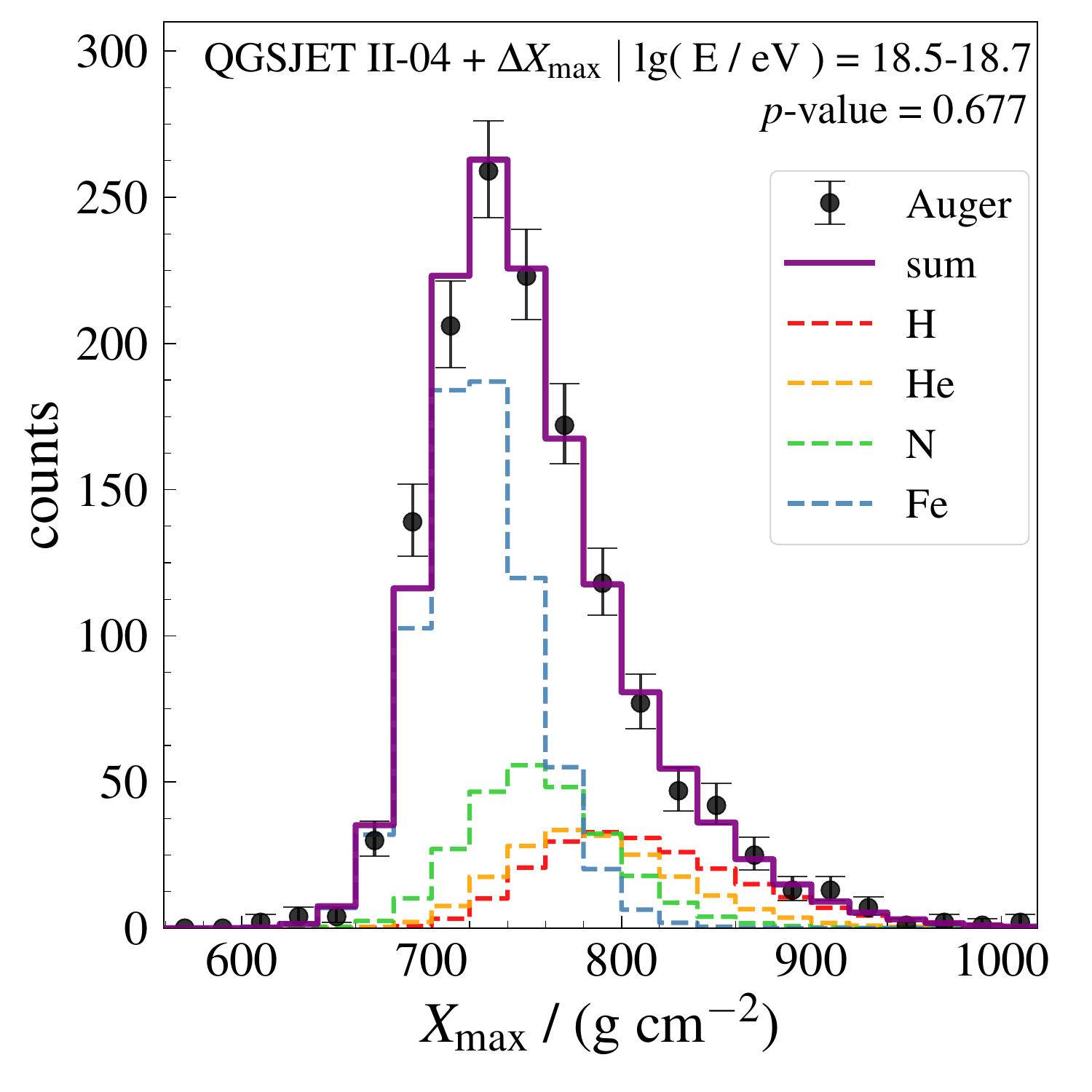} \\
        \includegraphics[width=0.24\textwidth]{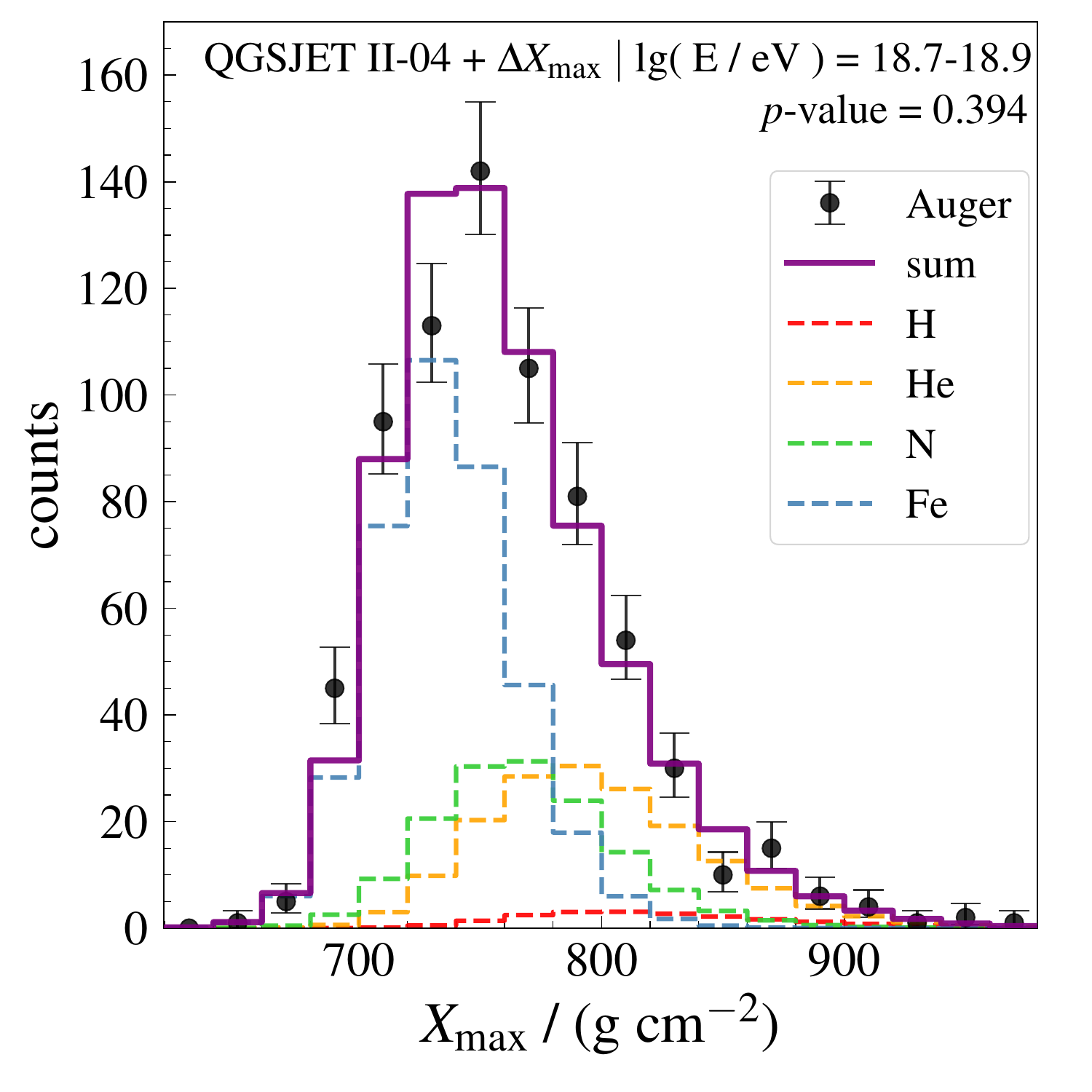} &
        \includegraphics[width=0.24\textwidth]{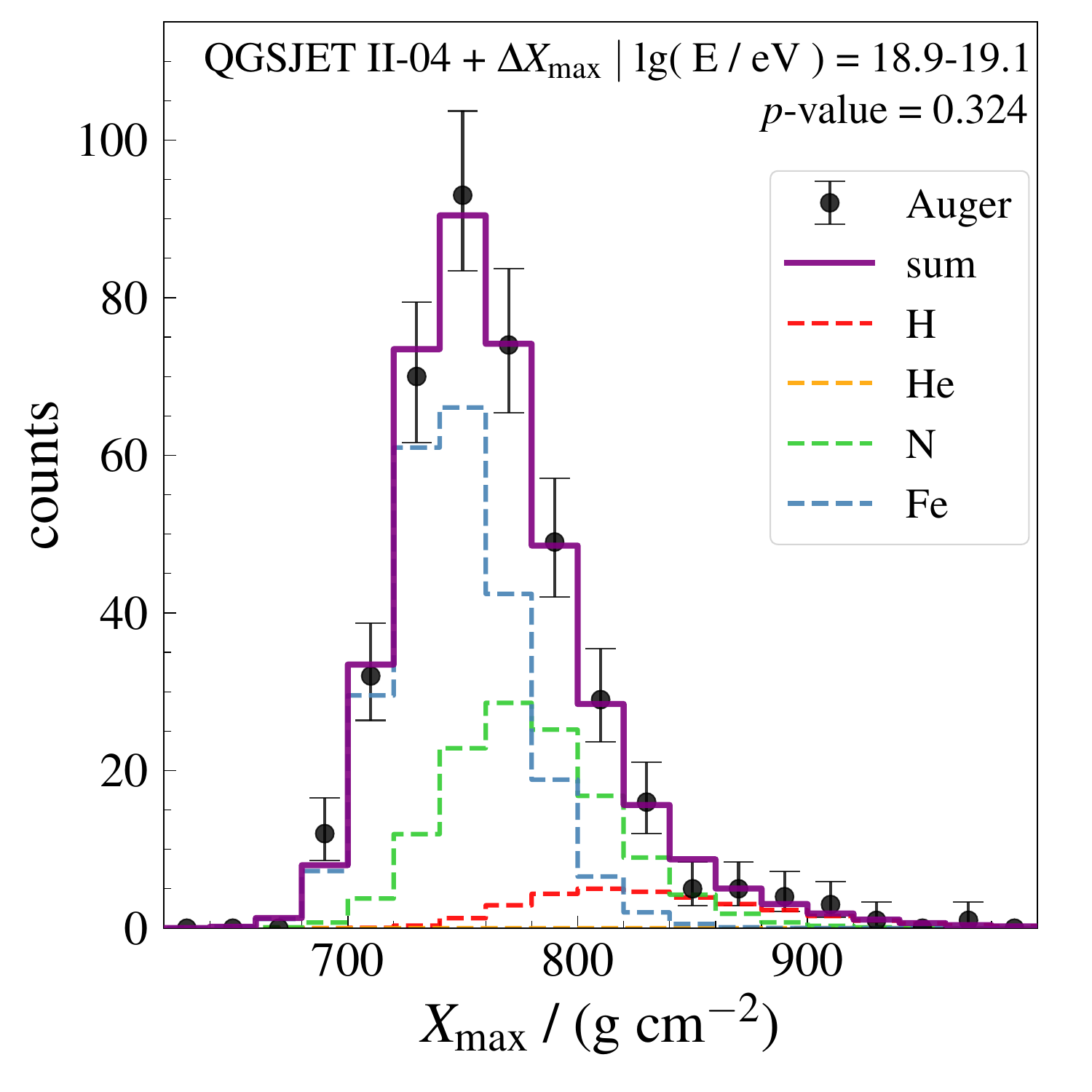} &
        \includegraphics[width=0.24\textwidth]{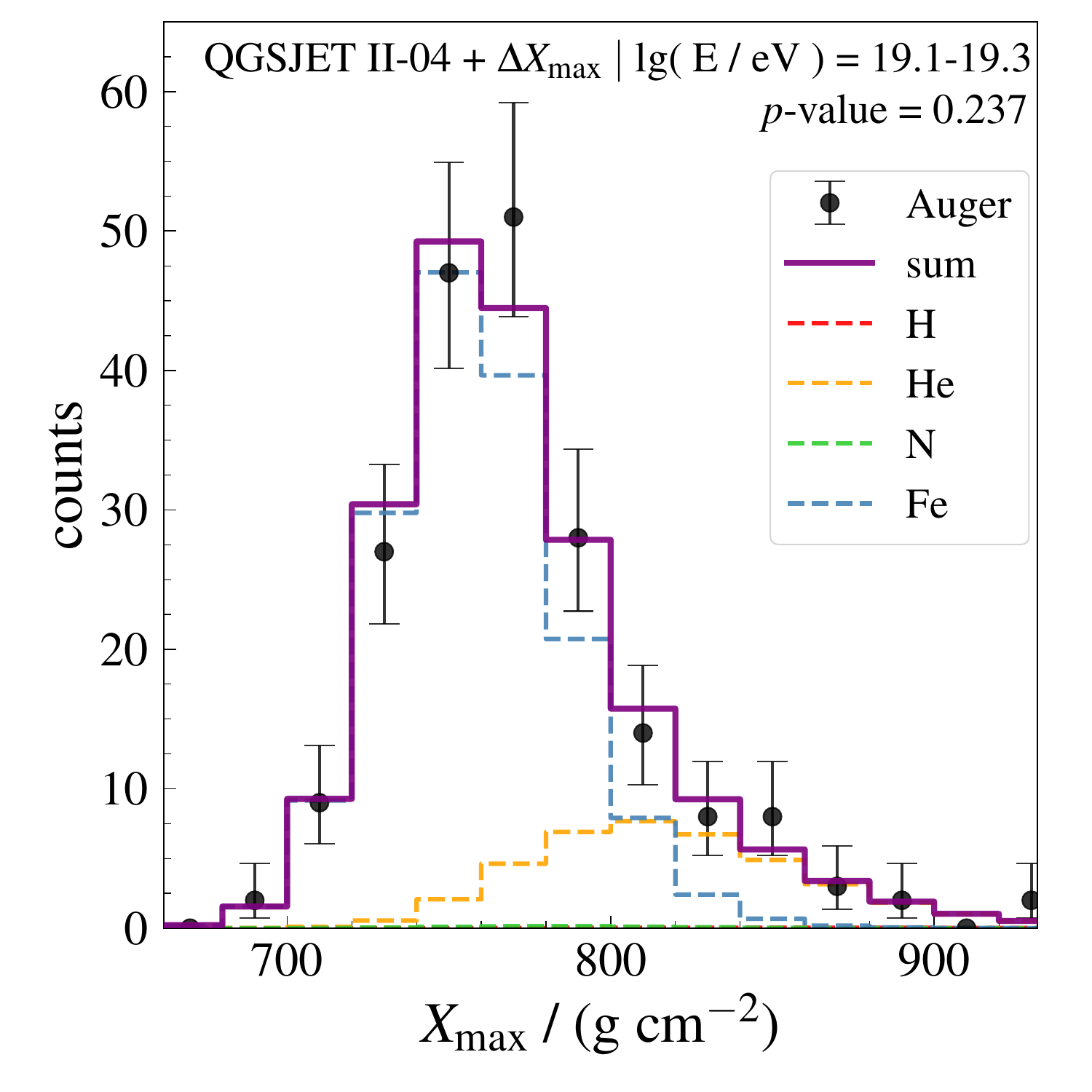} &
        \includegraphics[width=0.24\textwidth]{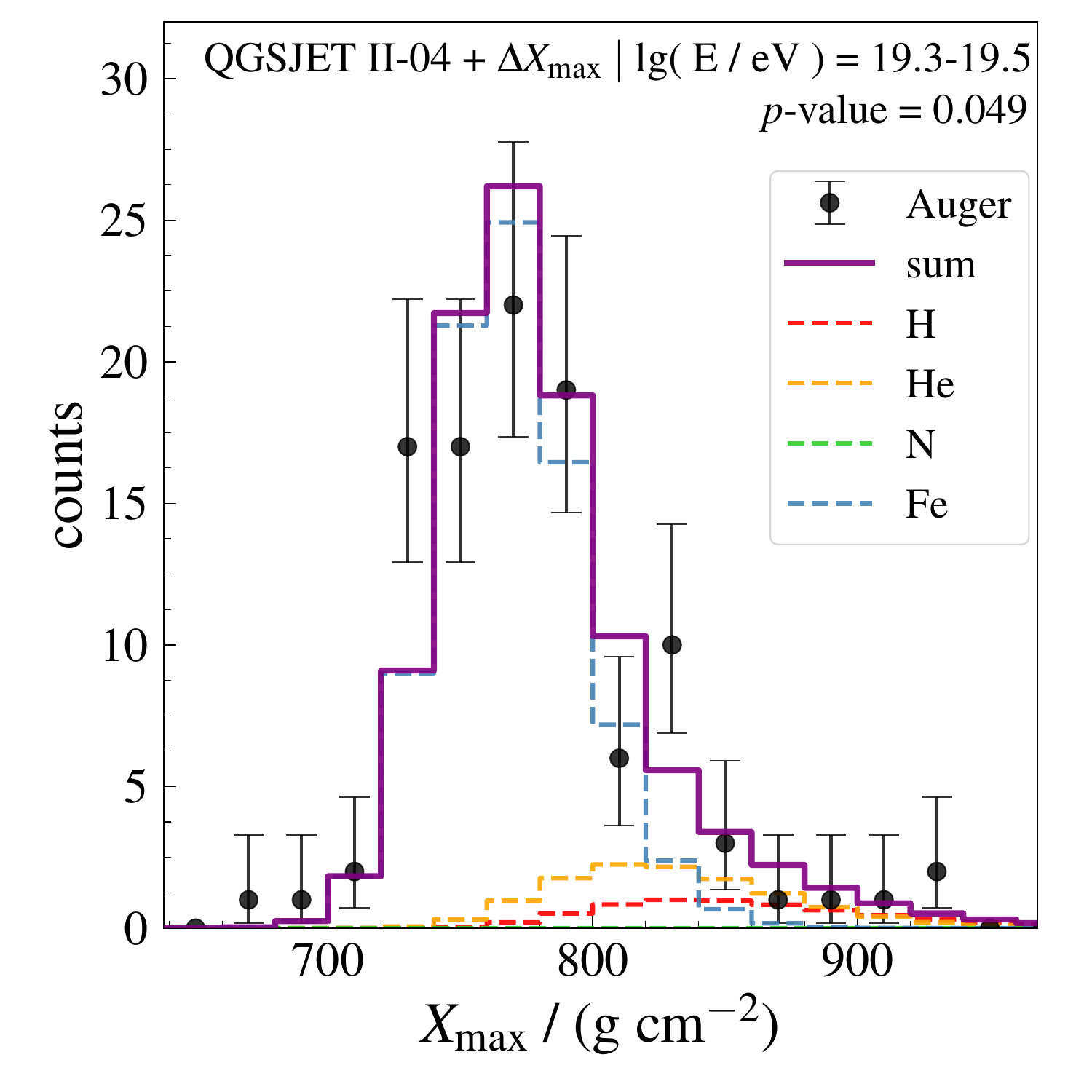} \\
        \includegraphics[width=0.24\textwidth]{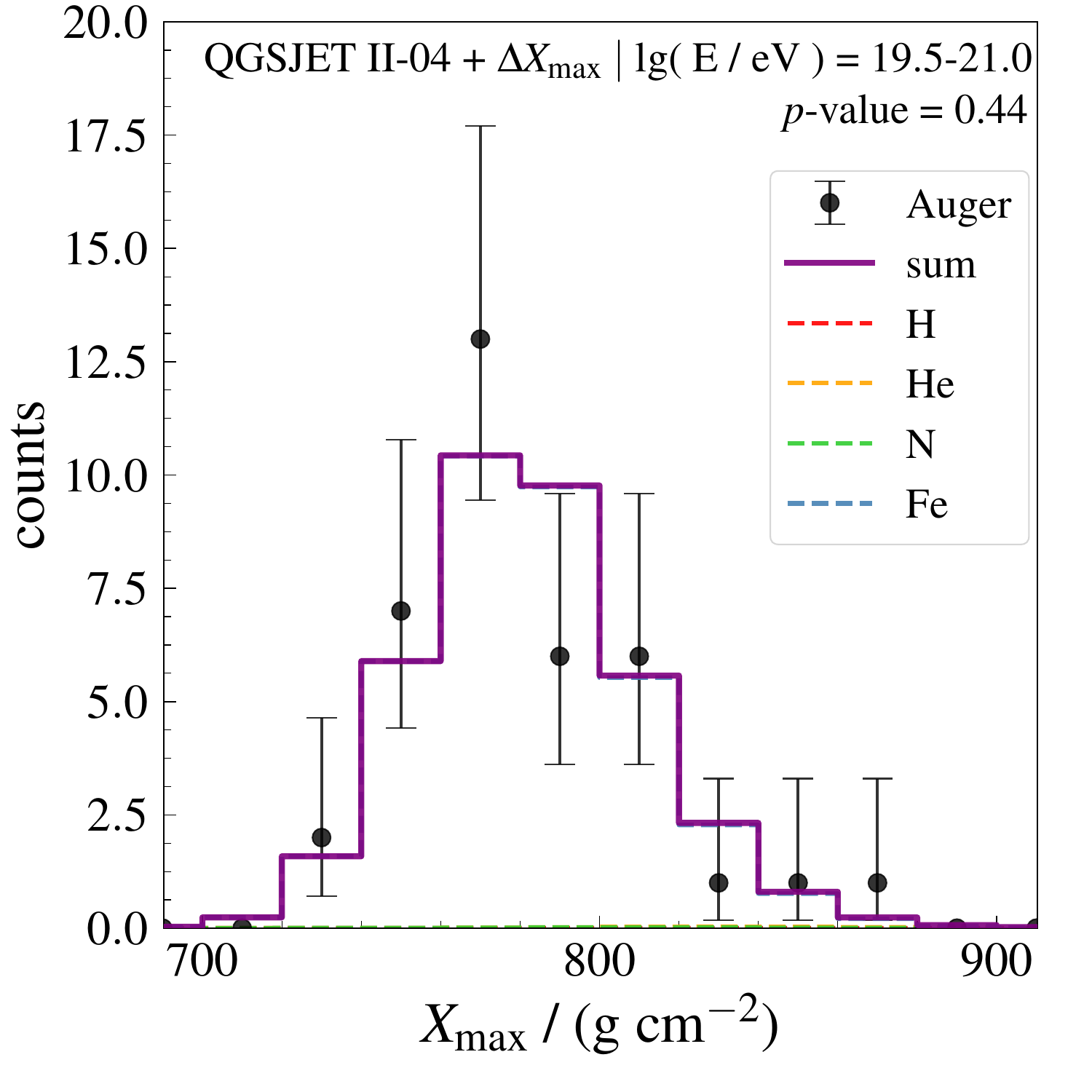} & & &
    \end{tabular}
    \caption{Same as for Fig.~\ref{fig:Xmax-distr-sibyll23d}, but for model \qgsii$+\DeltaXmax$.}
    \label{fig:Xmax-distr-qgsjetII04}
\end{figure}

We plot the Auger \Xmax distributions \citep{Auger-LongXmaxPaper} together with the simulated prediction and individual contributions of the four primaries in Fig.~\ref{fig:Xmax-distr-sibyll23d} for \sib{2.3d}$+\DeltaXmax$ and in Fig.~\ref{fig:Xmax-distr-qgsjetII04} for \qgsii$+\DeltaXmax$. For each hadronic interaction model, the energy binning intervals above $10^{18.4}$ eV were defined by merging adjacent bins based on the observed fluctuations in the fractions fitted with a step size of $\lg(E/ \eV)$ = 0.1.

\section{Arrival directions}
\label{app:Dipole}
\begin{figure}
    \centering
    \includegraphics[width=0.5\linewidth]{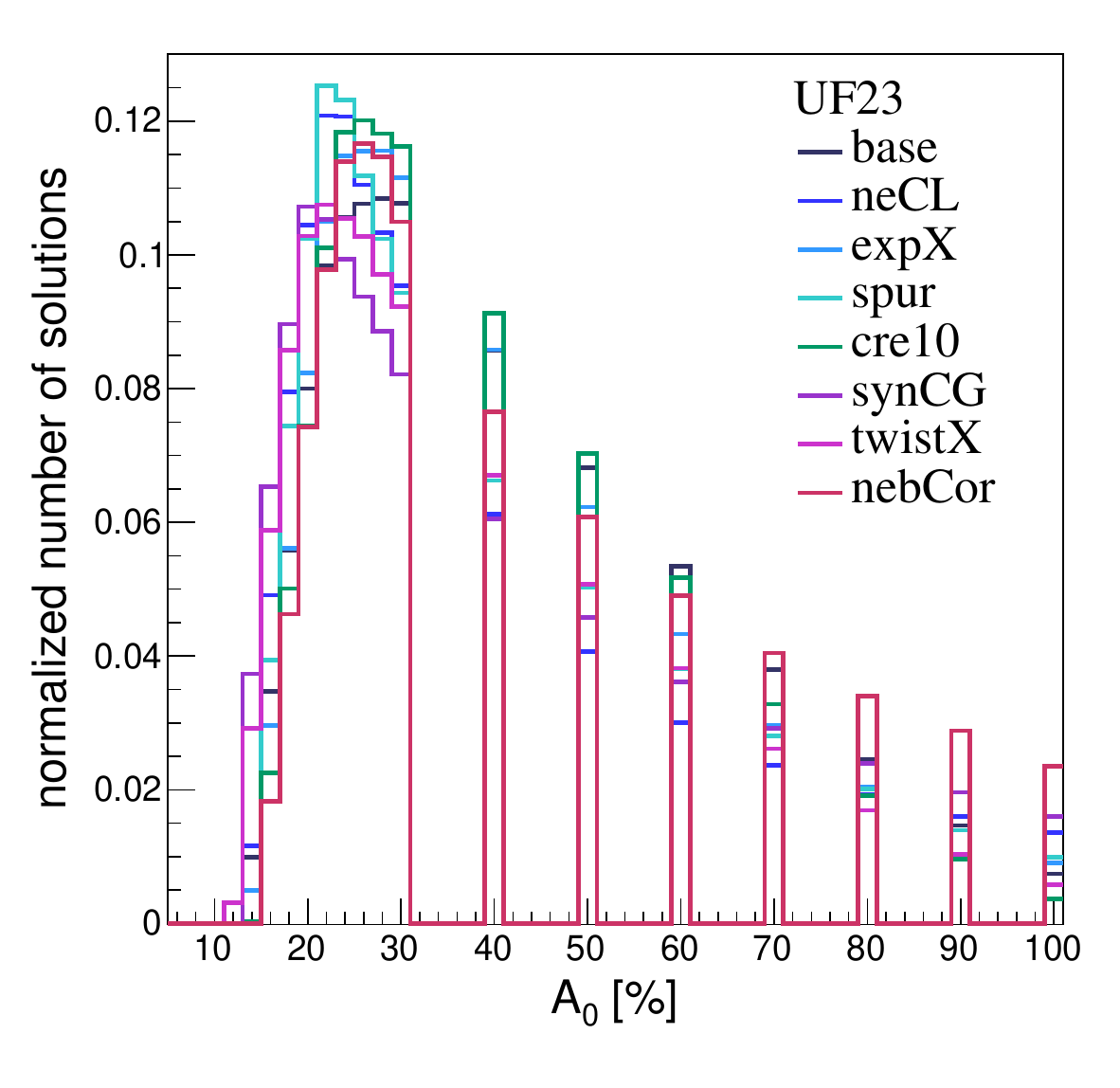}
    \includegraphics[width=0.7\linewidth]{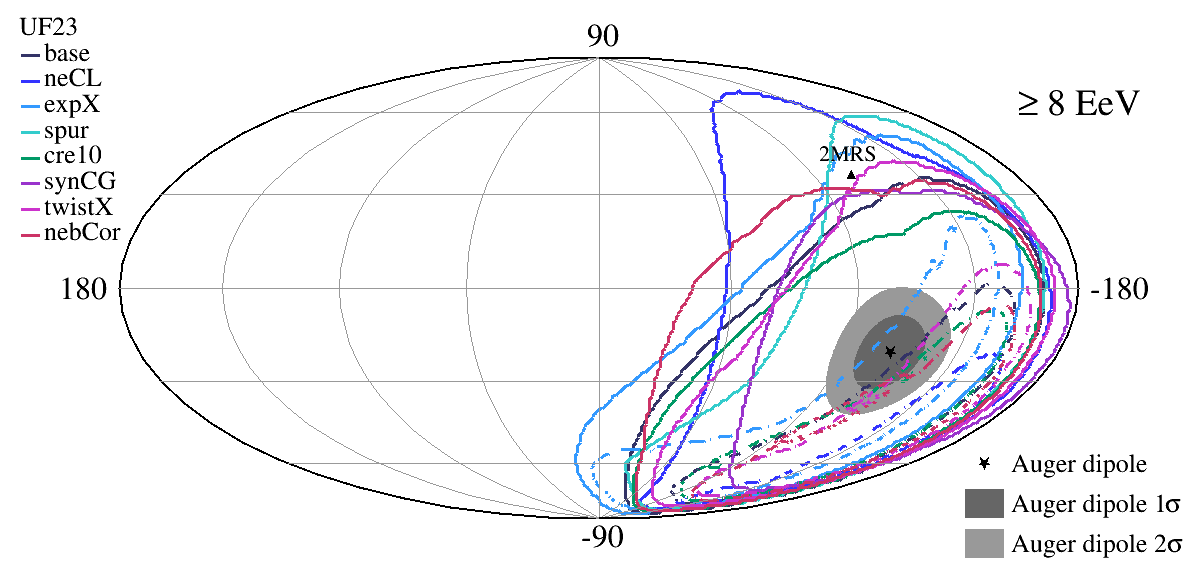}
    \includegraphics[width=0.7\linewidth]{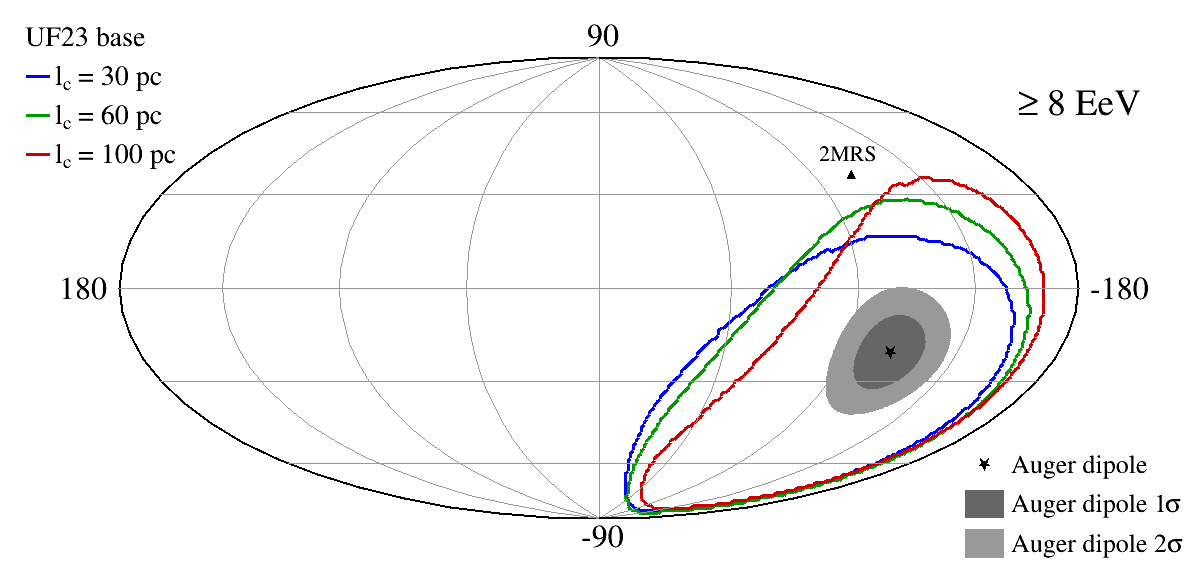}
    \caption{Top panel: The normalized distribution of the extragalactic amplitudes of the identified solutions from the top panel of Figure~\ref{fig:ExtragalacticDipoleUF23} for the individual models of the UF23 model of the GMF. \change{Middle} panel: Possible extragalactic directions of a dipole, compatible at the $1\sigma$ (dashed line) and $2\sigma$ (solid line) levels with the Auger measurement for the eight individual UF23 models of the GMF, shown in Galactic coordinates, using the mass composition obtained for model \sib{2.3d}$+\DeltaXmax$. \change{Bottom panel: Possible extragalactic directions of a dipole at the $2\sigma$ level for the UF23 base model of the GMF and three different coherence lengths, shown in Galactic coordinates, using the mass composition obtained for the model \sib{2.3d}$+\DeltaXmax$.}}
    \label{fig:DipoleAmplitudes}
\end{figure}

The $1\sigma$ and $2\sigma$ regions of the identified possible extragalactic directions of the dipole for individual UF23 models of the GMF are depicted in the \change{middle} panel of Fig.~\ref{fig:DipoleAmplitudes} \change{while the bottom panel shows the identified possible extragalactic directions of the dipole for three diferent coherence lengths of the turbulent field for the case of the UF23 base model.}.

The normalized distribution of the extragalactic amplitudes of the identified possible solutions of the dipole from Section~\ref{sec:Dipole} is shown in the top panel of Fig.~\ref{fig:DipoleAmplitudes} for the individual UF23 models of the GMF. As the mass composition of cosmic rays in the scenario proposed in this work is dominated by heavy nuclei, the required extragalactic amplitude of the dipole is rather large ($A_0\geq12\%$) in order to obtain a dipole amplitude on Earth compatible with the measured one. We decided to include extragalactic amplitudes as high as $100\%$ for completeness. However, such high amplitudes can not be easily explained by the spacial distribution of many sources and could be only achieved if a small number of sources contributes to the cosmic-ray flux above $8$~EeV. \change{We note that the maximum trajectory length of the simulated particles is set to 500 kpc. However, even for iron nuclei, the vast majority of the simulated particles have much shorter trajectory lengths, typically between $\sim (20 - 100)$ kpc.}

\begin{figure}
    \centering
    \includegraphics[width=0.75\linewidth]
    {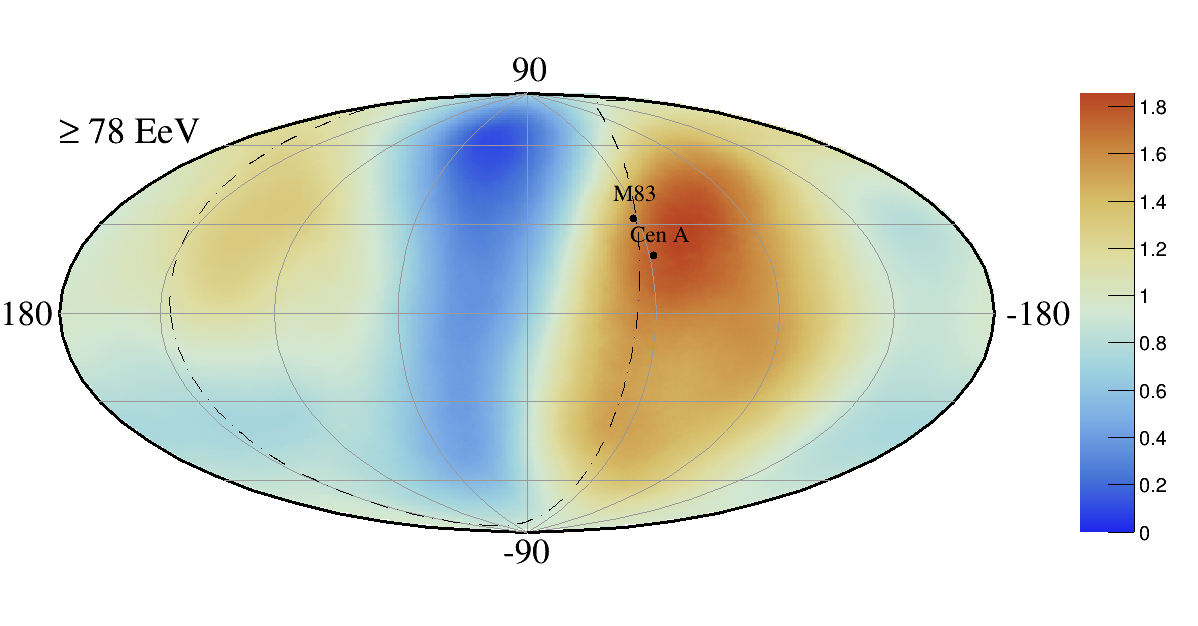}
    \caption{\change{The ratio of the 100 most energetic backtracked Auger events to the expectation for backtracked isotropically distributed arrival directions on Earth, smoothed with a 25$^{\circ}$ top-hat function, shown in Galactic coordinates. The supergalactic plane is depicted by dashed-and-dotted line with indicated directions of selected groups of galaxies.}}
    \label{fig:backtrackingIsotropic}
\end{figure}

In Fig.~\ref{fig:backtrackingIsotropic}, we show the distribution of the backtracked directions at the edge of the Galaxy for the isotropically distributed arrival directions on Earth with the same energies as the 100 most energetic events seen by the Pierre Auger Observatory \citep{AugerMostEnergeticEvents}. 
The geometrical exposure of the Pierre Auger Observatory is taken into account. 
The skymap shows similar features to the case of backtracking the real arrival directions of the 100 most energetic events, indicating that under the assumption of pure iron nuclei, most of the particles are coming from the galactic anticenter region.

\end{document}